\documentclass[11pt]{article}

\usepackage{amsmath}
\usepackage{amssymb}
\usepackage{amsthm}
\usepackage{color,soul}
\usepackage{tikz}
\usetikzlibrary{topaths,calc}
\usepackage{subfigure}
\usepackage{cite}
\usepackage{multirow}

\def\QED{\mbox{\rule[0pt]{1.5ex}{1.5ex}}}

\newtheorem{theorem}{Theorem}

\newtheorem{lemma}{Lemma}

\newtheorem{remark}{Remark}

\newcommand\blfootnote[1]{%
  \begingroup
  \renewcommand\thefootnote{}\footnote{#1}%
  \addtocounter{footnote}{-1}%
  \endgroup
}

\title{MDS Variable Generation and \\
Secure Summation with User Selection}
\author{\normalsize Yizhou Zhao and Hua Sun}
\date{}

\begin{document}
\maketitle
\blfootnote{
Yizhou Zhao (email: yizhouzhao@my.unt.edu) and Hua Sun (email: hua.sun@unt.edu) are with the Department of Electrical Engineering at the University of North Texas.}

\allowdisplaybreaks
\begin{abstract}
A collection of $K$ random variables are called $(K,n)$-MDS if any $n$ of the $K$ variables are independent and determine all remaining 
variables. In the MDS variable generation problem, 
$K$ users wish to generate variables that are $(K,n)$-MDS using a randomness variable owned by each user. We show that to generate $1$ bit of $(K,n)$-MDS variables for each $n \in \{1,2,\cdots, K\}$, the minimum size of the randomness variable at each user is $1 + 1/2 + \cdots + 1/K$ bits.

An intimately related problem is secure summation with user selection, where a server may select an arbitrary subset of $K$ users and securely compute the sum of the inputs of the selected users. We show that to compute $1$ bit of an arbitrarily chosen sum securely, the minimum size of the key held by each user is $1 + 1/2 + \cdots + 1/(K-1)$ bits, whose achievability uses the generation of $(K,n)$-MDS variables for $n \in \{1,2,\cdots,K-1\}$.

\end{abstract}

\newpage
\section{Introduction}
Maximum distance separable (MDS) codes are one of the most fascinating classes of codes in coding theory (see Chapter 11 of \cite{MacWilliams_Sloane}), with a wide array of applications ranging from storage systems \cite{Blaum_Bruck_Vardy, Dimakis_survey, VJ_Survey}, private information retrieval \cite{Banawan_Ulukus, FREIJ_HOLLANTI, Sun_Jafar_MDSTPIR, Zhou_Tian_Sun_Liu, Sun_Tian_MDS}, coded computation \cite{Kannan_Speeding, Dutta_Cadambe_Grover, Li_Salman} to secret sharing \cite{McEliece_Sarwate, Beimel_Survey} and secure multiparty computation \cite{BGW, CCD, cramer_damgard_nielsen_2015}. In this work, we take a Shannon theoretic view to study how to efficiently generate random variables that have the MDS property.

A collection of $K$ random variables $Z_1^n, \cdots, Z_K^n$ are said to be $(K,n)$-MDS 
if any $n$ of them are independent and uniquely determine the remaining variables (see Table \ref{table1} for an example). Consider $K$ users, where User $k \in \{1,2,\cdots, K\}$ holds a random variable $Z_k$. From $Z_k$, each user wishes to generate random variables $Z_k^1, \cdots, Z_k^K$ such that $Z_1^n, \cdots, Z_K^n$ are $(K,n)$-MDS. The question we explore is - to generate 1 bit of $Z_k^n$ for each $n \in \{1, 2, \cdots, K\}$, how many bits of the source $Z_k$ are required?

\begin{table}[!ht]
    \centering
        \begin{tikzpicture}[remember picture, overlay]
        \node (n1) at (10.3,-1.65) {};
        \node (Z5) at ($(n1)+(3.8,0.8)$) {};
        \draw[color=black]  ($(n1)$)-- ($(n1)+(3.5,0)$) -- ($(n1)+(3.5,3)$) -- ($(n1)+(0,3)$) -- ($(n1)$);
        \draw[color=black] ($(n1)+(3.5,1)$) -- ($(Z5)$);
        \node at ($(Z5)-(0.1,0)$) [right]{$Z_5$};
        
        \node (n2) at (10.4,-0.05) {};
        \node (Z5^3) at ($(n2)+(3.7,0.5)$) {};
        \draw[color=black]  ($(n2)$)-- ($(n2)+(3,0)$) -- ($(n2)+(3,0.5)$) -- ($(n2)+(0,0.5)$) -- ($(n2)$);
        \draw[color=black] ($(n2)+(3,0.25)$) -- ($(Z5^3)$);
        \node at ($(Z5^3)-(0.1,0)$) [right]{$Z_5^3$};
    \end{tikzpicture}
    \begin{tabular}{cllllll}
    \hline
         & User 1 & User 2 & User 3 & User 4 & User 5 & $\cdots$ \\ \hline
        $(K,1)$-MDS & $A_1$ & $A_1$ & $A_1$ & $A_1$ & $A_1$ & \\ 
        $(K,2)$-MDS & $A_2$ & $B_2$ & $A_2+B_2$ & $A_2+2B_2$ &  $A_2+3B_2$ \\ 
        $(K,3)$-MDS & $A_3$ & $B_3$ & $C_3$ & $A_3+B_3+C_3$ &  $A_3+2B_3+3C_3$\\
        $(K,4)$-MDS & $A_4$ & $B_4$ & $C_4$ & $D_4$ & $A_4+B_4+C_4+D_4$ \\
        \vdots &  &  &  &  &  \\ 
        $(K,K)$-MDS & $A_K$ & $B_K$ & $C_K$ & $D_K$ & $E_K$ & $\cdots$ \\ \hline
    \end{tabular}
    \caption{An example of MDS variables. 
    $A_i, B_j, \cdots$ are uniform and from a prime field, e.g., $\mathbb{F}_5$.}
    \label{table1}
\end{table}

From Table~\ref{table1}, we see that $K$ bits are sufficient for $Z_k$, when $Z_k^n$ are independent for each $n$. Interestingly, we show that if the correlation among $Z_k^n$ is optimally exploited, the size of $Z_k$ (normalized by the size of $Z_k^n$) can be reduced to $1 + 1/2 + \cdots + 1/K$, i.e., the harmonic number, which is roughly $\ln K$. Furthermore, information theoretic converse is provided to prove that this is absolutely minimum.

\begin{figure}[h]
\centering
\begin{tikzpicture}
    \node (u1) at (-5,0) {};
    \node (u2) at (-2.5,0) {};
    \node (u3) at (0,0) {};
    \node (u4) at (3.5,0) {};
    \node (u5) at (6,0.5) {$\cdots$};
    \node (server0) at (0,3) {};
    \node (server) at (-0.5,3) {};
    \node (server2) at ($(server)+(0.1,-0.2)$) {};
    \node (server3) at ($(server)+(0.1,0.1)$) {};
    \node (server4) at ($(server)+(0.1,0.4)$) {};
    
    \filldraw ($(u1) + (-0.5,0)$)
    to[out=90,in=180] ($(u1) + (0,0.5)$)
    to[out=0,in=90] ($(u1) + (0.5,0)$);
    \fill ($(u1) + (0,0.8)$) circle(0.3);
    \filldraw ($(u2) + (-0.5,0)$)
    to[out=90,in=180] ($(u2) + (0,0.5)$)
    to[out=0,in=90] ($(u2) + (0.5,0)$);
    \fill ($(u2) + (0,0.8)$) circle(0.3);
    \filldraw ($(u3) + (-0.5,0)$)
    to[out=90,in=180] ($(u3) + (0,0.5)$)
    to[out=0,in=90] ($(u3) + (0.5,0)$);
    \fill ($(u3) + (0,0.8)$) circle(0.3);
    \filldraw ($(u4) + (-0.5,0)$)
    to[out=90,in=180] ($(u4) + (0,0.5)$)
    to[out=0,in=90] ($(u4) + (0.5,0)$);
    \fill ($(u4) + (0,0.8)$) circle(0.3);
    
    \filldraw ($(server)+(0,-0.6+0.2+0.1)$) rectangle ($(server)+(1,0.7)$);
    \foreach \v in {2,...,4} {
        \filldraw [white] (server\v) rectangle ($(server\v)+(0.8,0.2)$);
        \filldraw ($(server\v)+(0.3,0.08)$) rectangle ($(server\v)+(0.75,0.12)$);
        \fill ($(server\v)+(0.15,0.1)$) circle (0.05);
        }
        
    \draw [rounded corners,->, line width=1pt,shorten >=16pt]($(u1) + (0,1.2)$) -- ($(u1) + (0,1.8)$)
    -- (server0);
    \draw [rounded corners,->, line width=1pt,shorten >=7pt]($(u3) + (0,1.2)$) -- ($(u3) + (0,1.8)$)
    -- (server0);
    \draw [rounded corners,->, line width=1pt,shorten >=16pt]($(u4) + (0,1.2)$) -- ($(u4) + (0,1.8)$)
    -- (server0);
    
    \node at ($(u1) + (0.5,0.15)$) [right]{User $1$};
    \node at ($(u2) + (0.5,0.15)$) [right]{User $2$};
    \node at ($(u3) + (0.5,0.15)$) [right]{User $3$};
    \node at ($(u4) + (0.5,0.15)$) [right]{User $4$};
    \node at ($(server0)+(-0.5,0.5)$) [left]{Server};
    
    \node at ($(u1) + (0,1.4)$) [right]{$W_1+A_2$};
    \node at ($(u2) + (0,1.4)$) [right]{};
    \node at ($(u3) + (0,1.4)$) [right]{$W_3-2\left(A_2+B_2\right)$};
    \node at ($(u4) + (0,1.4)$) [right]{$W_4+\left(A_2+2B_2\right)$};
    
    \node at ($(server0)+(1,0.5)$) [right]{only learn};
    \node at ($(server0)+(1,0)$) [right]{$W_1+W_3+W_4$};
\end{tikzpicture}
\caption{An example of secure summation, where the server selects User 1, User 3, and User 4 and securely computes the sum of their inputs $W_1+W_3+W_4$ using $(K,2)$-MDS key variables.}
\label{fig1}
\end{figure}
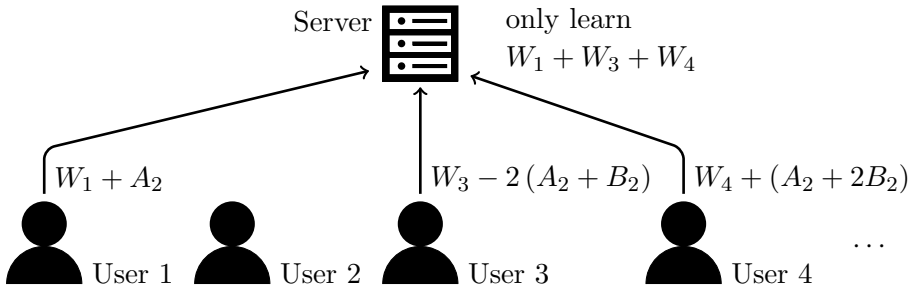

As an immediate application, we may use the generated MDS variables to the secure summation problem with user selection. In secure summation \cite{Zhao_Sun_Sum}, a server wishes to learn and only learn the sum of the inputs of a number of users. The problem of secure summation is motivated by the privacy need in aggregating the information from distributed users in federated learning \cite{aggregation, aggregation_log, aggregation_turbo, aggregation_fast, Zhao_Sun_Aggregate, aggregation_light, aggregation_swift, WSJC_Groupwise, aggregation_rosnes, aggregation_jun}. With user selection, the server may select an arbitrary subset of the $K$ users and securely compute their input sum.
Arbitrary user (client) selection (sampling, participation) is a common technique in federated learning \cite{cho2020client, mohamed2021privacy, zhao2021adaptive, fraboni2022general, wang2022unified}, which also gives rise to an interesting theoretical question on the randomness consumption, as we discuss next. 

From Figure~\ref{fig1}, we see that $(K,n)$-MDS variables may be used as the keys to securely compute the sum of $n+1$ selected users. As a result, MDS variable generation for $n \in \{1,2,\cdots, K-1\}$ can be applied to $K$-user secure summation with arbitrary user selection, i.e., each user holds a key of size $1+1/2+\cdots+1/(K-1)$ bits for each input bit. In addition, we show that such a key size is information theoretically optimal (minimum).

\section{Problem Statement}
In this section, we present the system model of the two problems that we study in this work - MDS variable generation and secure summation with user selection. 
\subsection{MDS Variable Generation
}

Consider $K$ users, where User $k \in \{1,2,\cdots, K\} \triangleq [K]$ holds a random variable $Z_k$ of size $L_Z$ bits. From $Z_k$, User $k \in [K]$ wishes to generate $K$ random variables, $(Z_k^1, \cdots, Z_k^K) = (Z_k^n)_{n\in [K]} \triangleq Z_k^{\leq K}
$, where each $Z_k^n$ has entropy $L$ bits.
\begin{eqnarray}
&& H(Z_k^1, \cdots, Z_k^K | Z_k) = 0,
\label{h1}\\
&& H(Z_k^n) = L, \forall n, k \in [K]. \label{h2}
\end{eqnarray}

Further, the variables $Z_1^n, Z_2^n, \cdots, Z_K^n$ are required to satisfy the following $(K,n)$-MDS property.
\begin{eqnarray}\label{generic}
H \left(\left(Z_k^n\right)_{k \in \mathcal{U}}\right) = \min\left(|\mathcal{U}|,n\right) \times L, ~\forall \mathcal{U} \subset [K]. 
\end{eqnarray}
In words, among $Z_1^n, Z_2^n, \cdots, Z_K^n$, any $n$ variables are independent and determine the remaining $K-n$ variables. 

The generation efficiency is measured by the rate $R_Z$, defined as follows.
\begin{eqnarray}
R_Z \triangleq \frac{L_Z}{L} \label{rate}
\end{eqnarray}
which characterizes the number of bits each user holds for each bit of the MDS variables generated.
A rate value $R_Z$ is said to be achievable
if there exists an MDS variable generation scheme (i.e., a design of variables $\left(Z_k^{n}\right)_{n,k \in [K]}$), for which constraints  (\ref{h1}), (\ref{h2}), (\ref{generic}) are satisfied, and the rate is no greater than $R_Z$. The infimum of achievable $R_Z$ values is called the optimal rate $R_Z^*$.

\subsection{Secure Summation with User Selection}
Consider $K$ users, where User $k \in [K]$ holds an input $W_k$ and a key $Z_k$. Each independent input $W_k$ is an $L \times 1$ vector and the $L$ elements are i.i.d. uniform symbols from the finite field $\mathbb{F}_q$. Each key $Z_k$ is an $L_Z \times 1$ vector over $\mathbb{F}_q$.  $(W_k)_{k \in [K]}$ is independent of $(Z_k)_{k \in [K]}$.
\begin{eqnarray}
    && H\left(\left(W_k\right)_{k\in[K]}, \left(Z_k\right)_{k\in[K]}\right) = \sum_{k\in[K]} H(W_k) + H\left(\left(Z_k\right)_{k\in[K]} \right),
\label{input_ind}\\
&& H(W_k) = L \log_2 q ~\mbox{bits}, \forall k \in [K]. \label{input_size_L}
\end{eqnarray}

Consider a server, who may select an arbitrary set of users $\mathcal{U} \subset [K]$ and wish to securely compute $\sum_{k \in \mathcal{U}} W_k$.
To this end, User $k \in \mathcal{U}$ sends a message $X_k^{\mathcal{U}}$ to the server, where $X_k^{\mathcal{U}}$ is a function of $W_k,Z_k$ and consists of $L_X$ symbols from $\mathbb{F}_q$.
\begin{eqnarray}
H(X_k^{\mathcal{U}} \big| W_k, Z_k) = 0, \forall k \in \mathcal{U}. \label{message}
\end{eqnarray}
From the messages received from the selected users, the server must be able to decode the desired sum $\sum_{k\in\mathcal{U}} W_k$ while nothing more is revealed in the information theoretic sense.
\begin{eqnarray}
\mbox{[Correctness]} && H\left(\sum_{k\in\mathcal{U}} W_k \Bigg| \left(X_k^{\mathcal{U}}\right)_{k\in\mathcal{U}}\right) = 0,  \forall \mathcal{U} \subset [K]. \label{corr} \\
\mbox{[Security]} && I\left(\left(W_k\right)_{k\in \mathcal{U}}; \left(X_k^{\mathcal{U}}\right)_{k\in \mathcal{U}} \Bigg| \sum_{k\in \mathcal{U}} W_k \right) = 0,  \forall \mathcal{U} \subset [K].
\label{sec}
\end{eqnarray}
The randomness consumption is measured by the key rate $R_Z$, defined as follows.
\begin{eqnarray}
R_Z \triangleq \frac{L_Z}{L}
\end{eqnarray}
which characterizes the number of symbols each key contains for each input symbol.
A rate value $R_Z$ is said to be achievable  
if there exists a secure summation scheme (i.e., a design of keys $\left(Z_k\right)_{k\in [K]}$ and messages $\left(X_k^{\mathcal{U}}\right)_{k\in \mathcal{U} \subset [K]}$), for which constraints (\ref{message}), (\ref{corr}), (\ref{sec}) are satisfied,
and the key rate is no greater than $R_Z$. The infimum of achievable $R_Z$ values is called the optimal key rate $R_Z^*$.

\section{Main Result}
Theorem \ref{theorem1} and Theorem \ref{theorem2} state the main result.
\begin{theorem}\label{theorem1}
For $K$-user MDS variable generation, the optimal rate is $R_Z^*=1 + 1/2 + \cdots + 1/K $.
\end{theorem}

The converse and achievability proof of Theorem \ref{theorem1} is presented in Section \ref{sec:mds_converse} and Section \ref{sec:mds_ach}, respectively.

\begin{theorem}\label{theorem2}
For $K$-user secure summation with arbitrary user selection, the optimal key rate is 
$R_Z^* = 1 + 1/2 + \cdots + 1/(K-1)$.
\end{theorem}

The converse and achievability proof of Theorem \ref{theorem2} is presented in Section \ref{sec:sum_converse} and Section \ref{sec:sum_ach}, respectively.

\section{Proof of Theorem \ref{theorem1}: Converse}\label{sec:mds_converse}
Before proceeding to the general proof, we first consider the setting where $K=3$ to illustrate the key ideas.

\subsection{Example: $K=3$ and $R_Z\geq 1+\frac{1}{2}+\frac{1}{3}$}
The converse proof has a recursive nature, where we consider 
the generation of $Z_k^{1}$, $Z_k^{\leq 2}$, and $Z_k^{\leq 3}$ successively and later steps rely on results obtained in previous steps.

{\it Step 1}: Consider $Z_k^1$. From the definition of MDS variables (\ref{generic}), we have 
\begin{eqnarray}
H(Z_k^1)= L, \forall k \in \{1,2,3\}.\label{ex1_con_step1}
\end{eqnarray}

{\it Step 2}: Consider $Z_k^{\leq 2}$.
\begin{eqnarray}
H\left(Z_1^{\leq 2}\right) + H\left(Z_2^{\leq 2}\right)
&=& H\left(Z_1^{\leq 2},Z_2^{\leq 2}\right) + I \left(Z_1^{\leq 2};Z_2^{\leq 2}\right)\label{ex1_con_step2_eq1}\\
&\geq& H\left(Z_1^{2},Z_2^{2}\right) + I \left(Z_1^{1};Z_2^{1}\right)\label{ex1_con_step2_eq2}\\
&\overset{(\ref{generic})}{=}& H\left(Z_1^{2},Z_2^{2}\right) + H\left(Z_1^1\right) \label{eq:ex1}\\
&\overset{(\ref{generic})(\ref{ex1_con_step1})}{=}& 2L + L 
~=~3L 
\label{ex1_con_step2}
\end{eqnarray}
where (\ref{eq:ex1}) follows from the definition of $(K,1)$-MDS variables, i.e., $Z_2^1$ is determined by $Z_1^1$, and in (\ref{ex1_con_step2}), the first term is due to definition of $(K,2)$-MDS variables and the second term follows from (\ref{ex1_con_step1}), i.e., the result from {\it Step 1} and we have reduced the problem from considering $Z_k^{\leq 2}$ to $Z_k^{1}$.

\begin{remark}\label{remark:ex1}
In the above derivation, 
one naively looking step (\ref{ex1_con_step2_eq2}) deserves highlighting. To obtain the first entropy term, we drop $Z_1^1,Z_2^1$ from $Z_1^{\leq 2},Z_2^{\leq 2}$ and this turns out to be tight because when we generate $(K,2)$-MDS variables $Z_k^{2}$, all entropy in $(K,1)$-MDS variables $Z_k^{1}$ is fully used (thus information wholly absorbed, see the achievable scheme in Section \ref{ex1_ach}). To obtain the second mutual information term, we drop $Z_1^2,Z_2^2$ because the two $(K,2)$-MDS variables are independent, leaving us with only $(K,1)$-MDS variables so that we may use results from Step 1. 
\end{remark}

Symmetrically, we can prove that (\ref{ex1_con_step2}) holds for any $2$ users, i.e., 
\begin{eqnarray}
H\left(Z_{i}^{\leq 2}\right) + H\left(Z_{j}^{\leq 2}\right) \geq 3L, \forall i, j \in \{1,2,3\}, i \neq j. \label{eq:ex12}
\end{eqnarray}

{\it Step 3}: Finally, consider $Z_k^{\leq 3}$. 
Denote the set of all permutations of $\{1,2,3\}$ as $\mathcal{S}_3 \triangleq \left\{\pi_i\right\}_{i \in 3!}$, where $\pi_i = (\pi_i(1), \pi_i(2), \pi_i(3))$
is a permutation of $\{1,2,3\}$. 
\begin{eqnarray}
&& 3! \times 3L_Z \notag\\
&\overset{(\ref{h1})}{\geq}& \sum_{\pi \in \mathcal{S}_3}  \Big[
H\left(Z_{\pi(1)}^{\leq 3}\right) + H\left(Z_{\pi(2)}^{\leq 3}\right) + H\left(Z_{\pi(3)}^{\leq 3}\right)  \Big] \\
&=& \sum_{\pi \in \mathcal{S}_3}  \Big[
 H\left(Z_{\pi(1)}^{\leq 3},Z_{\pi(2)}^{\leq 3},Z_{\pi(3)}^{\leq 3}\right) 
+ I\left(Z_{\pi(1)}^{\leq 3};Z_{\pi(2)}^{\leq 3}\right) 
+ I\left(Z_{\pi(3)}^{\leq 3};Z_{\pi(1)}^{\leq 3},Z_{\pi(2)}^{\leq 3}\right)  \Big] \label{eq:ex13}\\
&\geq& \sum_{\pi \in \mathcal{S}_3}  \Big[
H\left(Z_{\pi(1)}^{3},Z_{\pi(2)}^{3},Z_{\pi(3)}^{3}\right) 
+ I\left(Z_{\pi(1)}^{1};Z_{\pi(2)}^{1}\right) 
+ I\left(Z_{\pi(3)}^{\leq 2};Z_{\pi(1)}^{\leq 2},Z_{\pi(2)}^{\leq 2}\right)  \Big]
\label{ex1_con_step3_eq0}\\
&\overset{(\ref{generic})}{=}& 
3! H\left(Z_{1}^{3},Z_{2}^{3},Z_{3}^{3}\right) 
+ \sum_{\pi \in \mathcal{S}_3} 
H\left(Z_{\pi(1)}^{1}\right) + \sum_{\pi \in \mathcal{S}_3}  H\left(Z_{\pi(3)}^{\leq 2}\right) \label{ex1_con_step3_eq1} \\
&\overset{(\ref{generic})(\ref{ex1_con_step1})}{\geq}& 3!\times 3L + 3! \times L + \left[ H\left(Z_{1}^{\leq 2}\right) + H\left(Z_{2}^{\leq 2}\right) \right] + \left[ H\left(Z_{1}^{\leq 2}\right) + H\left(Z_{3}^{\leq 2}\right) \right] \notag\\
&& +~ \left[ H\left(Z_{2}^{\leq 2}\right) + H\left(Z_{3}^{\leq 2}\right) \right]
\label{ex1_con_step3_eq2}\\ 
&\overset{(\ref{eq:ex12})}{\geq}& 3!\times 3L + 3! \times L + 3 \times 3L 
\\
&\Rightarrow& ~ R_Z = L_Z/L \geq 1 + 1/2 + 1/3
\end{eqnarray}
where in (\ref{eq:ex13}), the identity $H(X)+H(Y)=H(X,Y)+I (X;Y)$ is used twice. 

\begin{remark}
Similar to Remark \ref{remark:ex1}, the key step is (\ref{ex1_con_step3_eq0}). For the first term, all entropy in $Z_k^{\leq 3}$ is preserved in $Z_k^3$; for the remaining two mutual information terms, we may drop the uncorrelated terms, after which they become the entropy terms in (\ref{ex1_con_step3_eq1}) due to the MDS property so that we may use results from Step 1 (i.e., $Z_k^1$) and Step 2 (i.e., $Z_k^{\leq 2}$).
\end{remark}

\subsection{General Proof: $R_Z \geq 1 + 1/2 + \cdots + 1/K$}
Let us start with two useful identities. The first identity, stated in the following lemma, transforms the sum of individual entropy terms to the sum of a joint entropy term and a number of mutual information terms.

\begin{lemma}\label{lemma_identity1}
For any random variables $Z_1, \cdots, Z_K$, we have
\begin{eqnarray}
&& H(Z_1) + H(Z_2) + \cdots + H(Z_K)\notag\\
&=& H(Z_1,Z_2,\cdots,Z_k) + I(Z_1;Z_2) + I(Z_3;Z_1,Z_2) + \cdots + I(Z_K;Z_1,Z_2,\cdots,Z_{K-1}). \label{lemma_identity1_eq}
\end{eqnarray}
\end{lemma}
{\it Proof:}
\begin{eqnarray}
&& \big[ H(Z_1) + H(Z_2) \big] + H(Z_3)+ \cdots + H(Z_K)\notag\\
&=& H(Z_1,Z_2) + I(Z_1;Z_2) + H(Z_3) + \cdots + H(Z_K)\\
&=& \big[ H(Z_1,Z_2) + H(Z_3)\big] + I(Z_1;Z_2) +  H(Z_4) + \cdots + H(Z_K)\\
&=& H(Z_1,Z_2,Z_3) + I(Z_3;Z_1,Z_2) + I(Z_1;Z_2) +  H(Z_4) + \cdots + H(Z_K)\\
&\vdots& \notag\\
&=&H(Z_1,Z_2,\cdots,Z_K) + I(Z_1;Z_2) + I(Z_3;Z_1,Z_2) + \cdots + I(Z_K;Z_1,Z_2,\cdots,Z_{K-1}).
\end{eqnarray}
\hfill\QED

The second identity, stated in the following lemma, transforms mutual information terms to joint entropy terms, for MDS variables.

\begin{lemma}\label{lemma_I_to_H}
For MDS variables $(Z_k^n)_{n,k\in[K]}$, we have 
\begin{eqnarray}\label{lemma_I_to_H_eq}
I\left(Z_k^{\leq n};\left(Z_u^{\leq n}\right)_{u \in \mathcal{U}}\right) = H\left(Z_k^{\leq n}\right), ~\forall \mathcal{U} \subset [K]\backslash \{k\}, |\mathcal{U}| = n.
\end{eqnarray}
\end{lemma}

{\it Proof:} The proof is immediate, by applying the definition of $(K,n)$-MDS variables in (\ref{generic}).
\begin{eqnarray}
I\left(Z_k^{\leq n};\left(Z_u^{\leq n}\right)_{u \in \mathcal{U}}\right) 
&=& H\left(Z_k^{\leq n}\right) + H\left(\left(Z_u^{\leq n}\right)_{u \in \mathcal{U}}\right) - H\left(\left(Z_u^{\leq n}\right)_{u \in \mathcal{U} \cup \{k\}}\right) \\
&\overset{(\ref{generic})}{=}& H\left(Z_k^{\leq n}\right).
\end{eqnarray}
\hfill\QED

We are now ready to recursively bound the entropy of any $n$ out of the $K$ MDS variables $Z_k^{\leq n}$. This result is stated in the following lemma.

\begin{lemma}\label{lemma_converse}
For MDS variables $(Z_k^n)_{n,k\in[K]}$, we have $\forall n \in [K]$
\begin{eqnarray}\label{lemma_converse_eq}
\frac{1}{n}\sum_{k \in \mathcal{U}} H\left(Z_k^{\leq n}\right) \geq \left(1+ \frac{1}{2} +\cdots + \frac{1}{n} \right) L, \forall \mathcal{U} \subset [K], |\mathcal{U}| = n.
\end{eqnarray}
\end{lemma}
{\it Proof:}
The proof is based on mathematical induction on $n$.

{\it Base case:} 
When $n=1$, (\ref{lemma_converse_eq}) becomes $H \left(Z_k^{1}\right) \geq L $, which follows directly from (\ref{generic}).

{\it Induction step:} Suppose (\ref{lemma_converse_eq}) holds for $n \in [M]$, $1 \leq M \leq K-1$, then we show that (\ref{lemma_converse_eq}) also holds for $n = M+1$.
Consider (\ref{lemma_converse_eq}) when $n = M+1$ and suppose  $\mathcal{U} = \{k_1,k_2,\cdots,k_{M+1}\} \subset [K]$. 
Denote the set of all permutations of $[M+1]$ as $\mathcal{S}_{M+1} = \left(\pi_i\right)_{i \in [(M+1)!]}$. 
\begingroup
\allowdisplaybreaks
\begin{eqnarray}
&& \frac{(M+1)!}{M+1}\sum_{k \in \mathcal{U}} H\left(Z_k^{\leq M+1}\right) 
\notag\\
&=& \frac{1}{(M+1)}\sum_{\pi \in \mathcal{S}_{M+1}} \left[H\left(Z_{k_{\pi (1)}}^{\leq M+1}\right)+H\left(Z_{k_{\pi (2)}}^{\leq M+1}\right)+ \cdots +H\left(Z_{k_{\pi (M+1)}}^{\leq M+1}\right)\right]\label{lemma_converse_eq1}\\
&\overset{(\ref{lemma_identity1_eq})}{=}& 
\frac{1}{(M+1)}\sum_{\pi \in \mathcal{S}_{M+1}} \left[H\left(Z_{k_{\pi (1)}}^{\leq M+1},Z_{k_{\pi (2)}}^{\leq M+1},\cdots,Z_{k_{\pi (M+1)}}^{\leq M+1}\right) + I\left(Z_{k_{\pi (1)}}^{\leq M+1};Z_{k_{\pi (2)}}^{\leq M+1}\right) \right.\notag\\
&&
~+ \left.I\left(Z_{k_{\pi (3)}}^{\leq M+1};Z_{k_{\pi (1)}}^{\leq M+1},Z_{k_{\pi (2)}}^{\leq M+1}\right) + \cdots + I\left(Z_{k_{\pi (M+1)}}^{\leq M+1};Z_{k_{\pi (1)}}^{\leq M+1},Z_{k_{\pi (2)}}^{\leq M+1},\cdots,Z_{k_{\pi (M)}}^{\leq M+1}\right)\right]\label{lemma_converse_eq2}\\
&\geq&
\frac{1}{(M+1)}\sum_{\pi \in \mathcal{S}_{M+1}} \left[H\left(Z_{k_{\pi (1)}}^{M+1},Z_{k_{\pi (2)}}^{ M+1},\cdots,Z_{k_{\pi (M+1)}}^{ M+1}\right) + I\left(Z_{k_{\pi (1)}}^{1};Z_{k_{\pi (2)}}^{1}\right) \right.\notag\\
&&
~+ \left.I\left(Z_{k_{\pi (3)}}^{\leq 2};Z_{k_{\pi (1)}}^{\leq 2},Z_{k_{\pi (2)}}^{\leq2}\right) + \cdots + I\left(Z_{k_{\pi (M+1)}}^{\leq M};Z_{k_{\pi (1)}}^{\leq M},Z_{k_{\pi (2)}}^{\leq M},\cdots,Z_{k_{\pi (M)}}^{\leq M}\right)\right]\label{lemma_converse_eq3}\\
&\overset{(\ref{generic})(\ref{lemma_I_to_H_eq})}{=}& 
\frac{1}{(M+1)}\sum_{\pi \in \mathcal{S}_{M+1}} \left[ (M+1)L + H\left(Z_{k_{\pi (1)}}^{1}\right) + H\left(Z_{k_{\pi (3)}}^{\leq 2}\right) + \cdots + H\left(Z_{k_{\pi (M+1)}}^{\leq M}\right)\right]\label{lemma_converse_eq4}\\
&=& 
\frac{1}{(M+1)}\left[
\sum_{\pi \in \mathcal{S}_{M+1}} (M+1)L 
+ 
\sum_{\pi \in \mathcal{S}_{M+1}} H\left(Z_{k_{\pi (1)}}^{1}\right) 
+
\frac{1}{2}\sum_{\pi \in \mathcal{S}_{M+1}} \left[H\left(Z_{k_{\pi (1)}}^{\leq 2}\right) +H\left(Z_{k_{\pi (2)}}^{\leq 2}\right)\right] \right.\notag\\
&&~+\left. \cdots 
+ \frac{1}{M}\sum_{\pi \in \mathcal{S}_{M+1}} \left[ H\left(Z_{k_{\pi (1)}}^{\leq M}\right) + H\left(Z_{k_{\pi (2)}}^{\leq M}\right) +\cdots + H\left(Z_{k_{\pi (M)}}^{\leq M}\right) \right]\right]\label{lemma_converse_eq5}\\
&\overset{}{\geq}& \frac{(M+1)!}{(M+1)} \left[(M+1)L + L + \left(1+\frac{1}{2}\right)L + \cdots + \left(1 + \frac{1}{2} + \cdots + \frac{1}{M}\right)L\right] ~\mbox{(Induction)}\label{lemma_converse_eq6}\\
&=& \frac{(M+1)!}{M+1}L \left[\left(M+1\right) + 1 \times M +\frac{1}{2} \times (M-1) + \frac{1}{3} \times (M-2) + \cdots + \frac{1}{M} \times 1\right] \\
&=&  \frac{(M+1)!}{M+1}L \bigg[\left(M+1\right) + (M+1)-1 +\frac{(M+1)-2}{2} + \frac{(M+1)-3}{3}  \notag\\
&&~~~~~~~~~~~~~~~~+ \cdots + \frac{(M+1)-M}{M}\bigg]\\
&=&  \frac{(M+1)!}{M+1}L \left[\left(M+1\right) + (M+1) +\frac{(M+1)}{2} + \frac{(M+1)}{3} + \cdots + \frac{(M+1)}{M} - M\right]\\
&=&  \frac{(M+1)!}{M+1}L \left[\left(M+1\right)  +\frac{(M+1)}{2} + \frac{(M+1)}{3} + \cdots + \frac{(M+1)}{M} + 1\right]\\
&=& (M+1)! \left(1  +\frac{1}{2} + \frac{1}{3} + \cdots + \frac{1}{M} + \frac{1}{M+1}\right)L \label{eq:end}
\end{eqnarray}
\endgroup
where in (\ref{lemma_converse_eq1}), we include all permutations of the users with indicies in $\mathcal{U}$ and (\ref{lemma_converse_eq2}) follows from Lemma \ref{lemma_identity1}.
In (\ref{lemma_converse_eq3}), we follow the insights in Remark \ref{remark:ex1} to drop terms, which cannot increase entropy or mutual information.  
In (\ref{lemma_converse_eq4}), 
we use Lemma \ref{lemma_I_to_H} and the definition of MDS variables (\ref{generic}).
In (\ref{lemma_converse_eq5}), we replace each term by averages using the property of all permutations, i.e., $\sum_{\pi \in \mathcal{S}_{M+1}} H\left(Z_{k_{\pi (i)}}^{\leq n}\right) = \sum_{\pi \in \mathcal{S}_{M+1}} H\left(Z_{k_{\pi (j)}}^{\leq n}\right), \forall i,j \in [M+1]$. 
In (\ref{lemma_converse_eq6}), we use the induction assumption that (\ref{lemma_converse_eq}) holds for $n \in [M]$. 

\hfill\QED

Equipped with Lemma \ref{lemma_converse}, the final converse proof of $R_Z$ follows immediately. Set $n=K$ in (\ref{lemma_converse_eq}), i.e., $\mathcal{U}= [K]$, then we have
\begin{eqnarray}
L_Z &\overset{(\ref{h1})}{\geq}& \frac{1}{K}\sum_{k \in [K]} H\left(Z_k^{\leq K}\right) \\
&\overset{(\ref{lemma_converse_eq})}{\geq}& \left(1+\frac{1}{2} +\cdots + \frac{1}{K} \right) L\\
\Rightarrow ~ R_Z = \frac{L_Z}{L} &\geq& 1+\frac{1}{2} +\cdots + \frac{1}{K}.
\end{eqnarray}

\section{Proof of Theorem \ref{theorem1}: Achievability}\label{sec:mds_ach}
The achievability proof is fairly straightforward. After setting up the dimensions following the insights from the converse proof, we only need to resort to random linear codes and random linear transformations. Let us start with an example of $K = 3$ to be familiar with the notations and then proceed to the general proof.

\subsection{Example: $K = 3$ and $R_Z = 1 + 1/2 + 1/3$} \label{ex1_ach}

We show that when $K = 3$, rate $R_Z = L_Z/L = 1 + 1/2 + 1/3  = 11/6$ is achievable. To this end, suppose $L = 6 \log_2 q$, i.e., each MDS variable $Z_k^n$ consists of $6$ symbols from $\mathbb{F}_q$ and $L_Z = 11 \log_2 q$, i.e., each source variable $Z_k$ consists of $11$ symbols from $\mathbb{F}_q$. Suppose the prime power field size $q > 72$.

{\it Step 1}: We describe the design of $Z_k$.  
We need 3 i.i.d. uniform $6\times1$ vectors over $\mathbb{F}_q$, denoted as $S^1,S^2,S^3$, then set
\begin{eqnarray}
Z_k = \left({\bf{H}}_k^{1}S^1, {\bf{H}}_k^{2} S^2,{\bf{H}}_k^{3} S^3\right), \forall k \in \{1,2,3\}
\end{eqnarray}
where ${{\bf{H}}_k^{1}} \in \mathbb{F}_q^{6 \times 6}, {{\bf{H}}_k^{2}} \in \mathbb{F}_q^{3 \times 6}, {{\bf{H}}_k^{3}} \in \mathbb{F}_q^{2 \times 6}$ need to satisfy some generic (full rank) properties (see Lemma \ref{lemma:full} for details). For now, it suffices to think of them as random matrices over a large field, which will work with high probability. Note that $Z_k$ has $6+3+2 = 11$ symbols, as desired.

{\it Step 2}: We describe the generation of MDS variables $Z_k^n$. 
For $(K,1)$-MDS variables $Z_k^1$, we set
\begin{eqnarray}
Z_k^1 = {\bf{H}}_k^{1}S^1 \label{eq:ach_zk1}
\end{eqnarray}
which has $6$ symbols.

For $(K,2)$-MDS variables $Z_k^2$, we set
\begin{eqnarray}
Z_k^2 = \left({\bf{V}}_k^{2\leftarrow 1}{\bf{H}}_k^{1}S^1,{\bf{H}}_k^{2}S^2\right) \label{eq:ach_zk2}
\end{eqnarray}
where ${\bf{V}}_k^{2\leftarrow 1} \in \mathbb{F}_q^{3 \times 6}$ transforms $(K,1)$-MDS variables to $(K,2)$-MDS variables (with maximum efficiency, see Remark \ref{remark:ex1} from the converse proof). Again, ${\bf{V}}_k^{2\leftarrow 1}$ need to satisfy some generic properties (stated later in Lemma \ref{lemma:full}), which hold with high probability over large fields. Note that $Z_k^2$ has $3+3 = 6$ symbols.

For $(K,3)$-MDS variables $Z_k^3$, we set
\begin{eqnarray}
&&Z_k^3 = \left({\bf{V}}_k^{3\leftarrow 1}{\bf{H}}_k^{1}S^1,
{\bf{V}}_k^{3\leftarrow 2}{\bf{H}}_k^{2}S^2,
{\bf{H}}_k^{3}S^3\right) \label{eq:ach_zk3}
\end{eqnarray}
where ${\bf{V}}_k^{3\leftarrow 1} \in \mathbb{F}_q^{2 \times 6}$ and ${\bf{V}}_k^{3\leftarrow 2} \in \mathbb{F}_q^{2 \times 3}$ transform $(K,1)$-MDS and $(K,2)$-MDS variables to $(K,3)$-MDS variables, respectively. The required conditions on ${\bf{V}}_k^{3\leftarrow 1}, {\bf{V}}_k^{3\leftarrow 2}$ will be stated later in Lemma \ref{lemma:full}, which are generic. Note that $Z_k^3$ has $2+2+2 = 6$ symbols.

{\it Step 3}: We specify the conditions on the matrices used in the code construction, ${\bf H}_k^n, {\bf V}_k^{n_2 \leftarrow n_1}$ such that MDS property (\ref{generic}) holds. For our proposed linear codes, it is straightforward to verify that we only need to guarantee $(\ref{generic})$ when\footnote{Our construction is based on linear transformations on uniform variables so that entropy terms boil down to rank terms. When $|\mathcal{U}| = n$, we will show that (\ref{generic}) is equivalent to requesting that certain square matrices ${\bf H}$ have full rank and $(Z_k^n)_{k \in \mathcal{U}}$ is invertible to $S^{\leq n}$ (refer to (\ref{eq:ach_h1}) to (\ref{eq:ach_z3})). As a result, when $|\mathcal{U}| < n$, (\ref{generic}) holds as it is associated with sub-matrices of ${\bf H}$, which must also have full rank; when $|\mathcal{U}| > n$, the additional $Z_k^n$ terms are a function of $S^{\leq n}$ thus contributing no more entropy.} $|\mathcal{U}| = n$.

For $(K,1)$-MDS variables $Z_k^1$, we require
\begin{eqnarray}
{\bf{H}}_k^{1} \in \mathbb{F}_q^{6 \times 6} ~\mbox{has full rank} \label{eq:ach_h1}
\end{eqnarray}
so that
\begin{eqnarray}
H(Z_k^1) \overset{(\ref{eq:ach_zk1})}{=} \mbox{rank}({\bf{H}}_k^{1}) \log_2 q \overset{(\ref{eq:ach_h1})}{=} 6 \log_2 q = L
\label{eq:ach_z1}
\end{eqnarray}
which follows from the uniformity of $S^1$ so that entropy of its linear transformation is specified by the rank of the transformation matrix ${\bf H}_k^{1}$.

For $(K,2)$-MDS variables $Z_k^2$, we require for any $\mathcal{U} = \{k_1, k_2\} \subset [K]$
\begin{eqnarray}
\left[
\begin{array}{c}
     {\bf{V}}_{k_1}^{2\leftarrow 1}{\bf{H}}_{k_1}^{1}\\
     {\bf{V}}_{k_2}^{2\leftarrow 1}{\bf{H}}_{k_2}^{1}
\end{array}
\right]_{6\times 6}
~\mbox{and}~
\left[
\begin{array}{c}
    {\bf{H}}_{k_1}^{2}\\
    {\bf{H}}_{k_2}^{2}
\end{array}
\right]_{6\times 6}
~\mbox{have full rank} 
\label{eq:ach_h2}
\end{eqnarray}
so that
\begin{eqnarray}
H(Z_{k_2}^2, Z_{k_3}^2) &\overset{(\ref{eq:ach_zk2})}{=}& H \left({\bf{V}}_{k_1}^{2\leftarrow 1}{\bf{H}}_{k_1}^{1}S^1,
{\bf{V}}_{k_2}^{2\leftarrow 1}{\bf{H}}_{k_2}^{1}S^1,
{\bf{H}}_{k_1}^{2}S^2,
{\bf{H}}_{k_2}^{2}S^2 \right) \\
&=& H\left( \left[
\begin{array}{c}
     {\bf{V}}_{k_1}^{2\leftarrow 1}{\bf{H}}_{k_1}^{1}\\
     {\bf{V}}_{k_2}^{2\leftarrow 1}{\bf{H}}_{k_2}^{1}
\end{array}
\right]  S^1 \right)
+ 
H\left( \left[
\begin{array}{c}
    {\bf{H}}_{k_1}^{2}\\
    {\bf{H}}_{k_2}^{2}
\end{array}
\right] S^2 \right) \label{eq:ach_t1} \\
&=& \left( \mbox{rank} \left( \left[
\begin{array}{c}
     {\bf{V}}_{k_1}^{2\leftarrow 1}{\bf{H}}_{k_1}^{1}\\
     {\bf{V}}_{k_2}^{2\leftarrow 1}{\bf{H}}_{k_2}^{1}
\end{array}
\right] \right) + \mbox{rank}
\left( \left[
\begin{array}{c}
    {\bf{H}}_{k_1}^{2}\\
    {\bf{H}}_{k_2}^{2}
\end{array}
\right] \right)
\right) \log_2 q\\
&\overset{(\ref{eq:ach_h2})}{=}& 12 \log_2 q = 2L
\label{eq:ach_z2}
\end{eqnarray}
where (\ref{eq:ach_t1}) follows from the independence of $S^1$ and $S^2$.

For $(K,3)$-MDS variables $Z_k^3$, we require
\begin{eqnarray}
\left[\begin{array}{c}
    {\bf{V}}_1^{3\leftarrow 1} {\bf{H}}_1^{1}\\
    {\bf{V}}_2^{3\leftarrow 1} {\bf{H}}_2^{1}\\
    {\bf{V}}_3^{3\leftarrow 1} {\bf{H}}_3^{1}
\end{array}
\right]_{6\times 6}, 
\left[\begin{array}{c}
    {\bf{V}}_1^{3\leftarrow 2} {\bf{H}}_1^{2}\\
    {\bf{V}}_2^{3\leftarrow 2} {\bf{H}}_2^{2}\\
    {\bf{V}}_3^{3\leftarrow 2} {\bf{H}}_3^{2}
\end{array}
\right]_{6\times 6},
~\mbox{and}
\left[\begin{array}{c}
    {\bf{H}}_1^{3}\\
    {\bf{H}}_2^{3}\\
    {\bf{H}}_3^{3}
\end{array}
\right]_{6\times 6}
~\mbox{have full rank}
\label{eq:ach_h3}
\end{eqnarray}
so that
\begin{eqnarray}
&& H(Z_{1}^3, Z_{2}^3, Z_3^3)\notag\\
&\overset{(\ref{eq:ach_zk3})}{=}& H \left({\bf{V}}_{1}^{3\leftarrow 1}{\bf{H}}_{1}^{1}S^1,
{\bf{V}}_{2}^{3\leftarrow 1}{\bf{H}}_{2}^{1}S^1,
{\bf{V}}_{3}^{3\leftarrow 1}{\bf{H}}_{3}^{1}S^1 \right) \notag\\
&& ~+ H \left({\bf{V}}_{1}^{3\leftarrow 2}{\bf{H}}_{1}^{2}S^2,
{\bf{V}}_{2}^{3\leftarrow 2}{\bf{H}}_{2}^{2}S^2,
{\bf{V}}_{3}^{3\leftarrow 2}{\bf{H}}_{3}^{2}S^2 \right) 
+ H \left( {\bf{H}}_{1}^{3}S^3,
{\bf{H}}_{2}^{3}S^3,
{\bf{H}}_{3}^{3}S^3 \right) \\
&=& \left( 
\mbox{rank}\left(
\left[\begin{array}{c}
    {\bf{V}}_1^{3\leftarrow 1} {\bf{H}}_1^{1}\\
    {\bf{V}}_2^{3\leftarrow 1} {\bf{H}}_2^{1}\\
    {\bf{V}}_3^{3\leftarrow 1} {\bf{H}}_3^{1}
\end{array}
\right] \right) + 
\mbox{rank}\left(
\left[\begin{array}{c}
    {\bf{V}}_1^{3\leftarrow 2} {\bf{H}}_1^{2}\\
    {\bf{V}}_2^{3\leftarrow 2} {\bf{H}}_2^{2}\\
    {\bf{V}}_3^{3\leftarrow 2} {\bf{H}}_3^{2}
\end{array}
\right] \right) +
~\mbox{rank}
\left(
\left[\begin{array}{c}
    {\bf{H}}_1^{3}\\
    {\bf{H}}_2^{3}\\
    {\bf{H}}_3^{3}
\end{array}
\right] \right)
\right) \log_2 q\\
&\overset{(\ref{eq:ach_h3})}{=}& 18 \log_2 q = 3L.
\label{eq:ach_z3}
\end{eqnarray}

{\it Step 4:} Finally, we show that there exist matrices ${\bf H}_k^n, {\bf V}_k^{n_2 \leftarrow n_1}$ that satisfy the required full rank conditions obtained in the previous step. This result is stated in the following lemma.
\begin{lemma}\label{lemma:full}
When $q > 72$, there exist  $\left( {\bf H}_k^n \right)_{k, n \in \{1,2,3\}}, \left( {\bf V}_k^{n_2 \leftarrow n_1} \right)_{k,n_1,n_2\in\{1,2,3\}, n_2 > n_1}$ such that (\ref{eq:ach_h1}), (\ref{eq:ach_h2}), (\ref{eq:ach_h3}) are satisfied. 
\end{lemma}

{\it Proof:} The existence proof is based on probabilistic arguments. Draw each element of the matrices ${\bf H}_k^n, {\bf V}_k^{n_2 \leftarrow n_1}$ independently and uniformly from $\mathbb{F}_q$. Denote the vector that contains all such elements as $\vec{v}$. View the determinant of each matrix in (\ref{eq:ach_h1}), (\ref{eq:ach_h2}), (\ref{eq:ach_h3}) as a polynomial in $\vec{v}$ and consider the product of all such polynomials, denoted by $f(\vec{v})$.
$f(\vec{v})$ is product of $\binom{3}{1} + 2\binom{3}{2} + 3\binom{3}{3} = 12 $ polynomials, each of which has degree at most $6$, so the degree of $f(\vec{v})$ is at most $12 \times 6 = 72$.

$f(\vec{v})$ is not the zero polynomial (proved later), so we can apply the Schwartz–Zippel lemma to obtain
\begin{eqnarray}
\Pr(f(\vec{v}) = 0) \leq 72/q < 1.
\end{eqnarray}
Therefore, there exists at least one assignment of ${\bf H}_k^n, {\bf V}_k^{n_2 \leftarrow n_1}$ so that all matrices in (\ref{eq:ach_h1}), (\ref{eq:ach_h2}), (\ref{eq:ach_h3}) have full rank and thus the generated variables are indeed MDS.

Lastly, we are left to prove that $f(\vec{v})$ is not identically zero. To this end, it suffices to consider each matrix in (\ref{eq:ach_h1}), (\ref{eq:ach_h2}), (\ref{eq:ach_h3}) and show that for each such matrix, there exists one realization of ${\bf H}_k^n, {\bf V}_k^{n_2 \leftarrow n_1}$ so that the matrix has full rank (and its determinant polynomial is not identically zero). This is proved next. A matrix that only involves ${\bf H}_k^n$ is trivial as we may set it as the identity matrix; a matrix that involves both ${\bf H}_k^n$ and ${\bf V}_k^{n_2 \leftarrow n_1}$ can be set as the identity matrix as well because
\begin{eqnarray}
(\ref{eq:ach_h2}):&&
\left[
\begin{array}{c}
    {\bf{V}}_{k_1}^{2 \leftarrow 1}{\bf{H}}_{k_1}^{1}\\
    {\bf{V}}_{k_2}^{2 \leftarrow 1}{\bf{H}}_{k_2}^{1}
\end{array}\right] = {\bf{I}}_6
~\Leftarrow~ {\bf{V}}_{k_1}^{2\leftarrow 1} = {\bf{V}}_{k_2}^{2\leftarrow 1} = \left[
\begin{array}{c c}
    {\bf{I}}_{3} &  {\bf{0}}_{3}
\end{array} \right] \\
&&~~~~~~~~~~~~~~~~~~~~~~~~~~~~~~~
{\bf{H}}_{k_1}^{1} = \left[
\begin{array}{cc}
     {\bf{I}}_{3} & {\bf{0}}_{3} \\
     {\bf{0}}_{3} & {\bf{0}}_{3}
\end{array}\right]
,~ {\bf{H}}_{k_2}^{1} = \left[\begin{array}{cc}
     {\bf{0}}_{3} & {\bf{I}}_{3} \\
     {\bf{0}}_{3} & {\bf{0}}_{3}
\end{array}\right]\\
(\ref{eq:ach_h3}):&&
\left[
\begin{array}{c}
    {\bf{V}}_1^{3\leftarrow 1}{\bf{H}}_1^{1}\\
    {\bf{V}}_2^{3\leftarrow 1}{\bf{H}}_2^{1}\\
    {\bf{V}}_3^{3\leftarrow 1}{\bf{H}}_3^{1}
\end{array}\right] = {\bf{I}}_6 
~\Leftarrow~
{\bf{V}}_{1}^{3\leftarrow 1}=
     {\bf{V}}_{2}^{3\leftarrow 1}=
     {\bf{V}}_{3}^{3\leftarrow 1} = 
     \left[\begin{array}{c c}
    {\bf{I}}_{2} &  {\bf{0}}_{2\times4}
\end{array} \right] \notag\\
&&
{\bf{H}}_{1}^{1} = \left[
\begin{array}{cc}
     {\bf{I}}_{2} & {\bf{0}}_{2\times 4} \\
     {\bf{0}}_{4\times 2} & {\bf{0}}_{4 \times 4}
\end{array}\right]
,~ 
{\bf{H}}_{2}^{1} = \left[
\begin{array}{ccc}
     {\bf{0}}_{2} &{\bf{I}}_{2} & {\bf{0}}_{2} \\
     {\bf{0}}_{4 \times 2} & {\bf{0}}_{4\times 2} & {\bf{0}}_{4\times 2}
\end{array}\right]
,~
{\bf{H}}_{3}^{1} = \left[
\begin{array}{ccc}
     {\bf{0}}_{2\times 4} &{\bf{I}}_{2} \\
     {\bf{0}}_{4\times 4} & {\bf{0}}_{4\times 2}
\end{array}\right];
\\
&&\left[
\begin{array}{c}
    {\bf{V}}_1^{3\leftarrow 2}{\bf{H}}_1^{2}\\
    {\bf{V}}_2^{3\leftarrow 2}{\bf{H}}_2^{2}\\
    {\bf{V}}_3^{3\leftarrow 2}{\bf{H}}_3^{2}
\end{array}\right] = {\bf{I}}_6 ~\Leftarrow~
     {\bf{V}}_{1}^{3\leftarrow 2}=
     {\bf{V}}_{2}^{3\leftarrow 2}=
     {\bf{V}}_{3}^{3\leftarrow 2} = 
     \left[
\begin{array}{c c}
    {\bf{I}}_{2} &  {\bf{0}}_{2\times1}
\end{array} \right],\notag\\
&&
    {\bf{H}}_{1}^{2} = \left[
\begin{array}{cc}
     {\bf{I}}_{2} & {\bf{0}}_{2\times 4} \\
     {\bf{0}}_{1\times 2} & {\bf{0}}_{1 \times 4}
\end{array}\right]
,~ 
{\bf{H}}_{2}^{2} = \left[
\begin{array}{ccc}
     {\bf{0}}_{2} &{\bf{I}}_{2} & {\bf{0}}_{2} \\
     {\bf{0}}_{2} & {\bf{0}}_{2} & {\bf{0}}_{2}
\end{array}\right]
,~
{\bf{H}}_{3}^{2} = \left[
\begin{array}{cc}
     {\bf{0}}_{2\times 4} &{\bf{I}}_{2} \\
     {\bf{0}}_{1\times 4} & {\bf{0}}_{1\times 2}
\end{array}\right]
\end{eqnarray}
where ${\bf I}_i$ is the $i \times i$ identity matrix and ${\bf 0}_i \hspace{0.07cm} ( {\bf 0}_{i \times j})$ is an $i\times i ~( i \times j)$ matrix wherein each element is zero.

\hfill\QED

\subsection{General Proof: Any $K$}
The general achievability proof of $R_Z = 1 + 1/2 + \cdots + 1/K$ is an immediate generalization of that of above example. Suppose $L = K! \log_2 q$ and the prime power field size\footnote{Similar to Shannon's original random coding proof to the achievability of channel capacity, our proof is existence based and no effort is devoted to minimizing the field size required.} $q > K! \sum_{n\in[K]} n \binom{K}{n}$.

{\it Step 1}: Design $Z_k$. Set  
\begin{eqnarray}
Z_k =  \left(\left({\bf{H}}_k^{n}S^n\right)_{n\in[K]}\right), \forall k \in [K] \label{eq:ach_zkg}
\end{eqnarray}
where $S^n, n \in [K]$ are $K$ i.i.d. uniform $K! \times1$ vectors over $\mathbb{F}_q$ and ${\bf{H}}_k^{n} \in \mathbb{F}_q^{\frac{K!}{n} \times K!}$. 
Note that $Z_k$ contains $L_Z/\log_2 q = \sum_{n \in [K]} K!/n$ symbols, so $R_Z = L_Z/L=\sum_{n\in [K]}1/n$, as desired. 

{\it Step 2}: Design $Z_k^n$. Set
\begin{eqnarray}
Z_k^n=\left(\left({\bf{V}}_k^{n\leftarrow m}{\bf{H}}_k^{m} S^m\right)_{m \in [n-1]},{\bf{H}}_k^{n}S^n\right),\forall m \in [K] \label{eq:ach_zkng}
\end{eqnarray}
where ${\bf{V}}_k^{n\leftarrow m} \in \mathbb{F}_q^{\frac{K!}{n} \times \frac{K!}{m}}$.

{\it Step 3}: Conditions on ${\bf H}_k^n, {\bf V}_k^{n_2 \leftarrow n_1}$ such that MDS property (\ref{generic}) holds. 
For $(K,n)$-MDS variables $Z_k^n$, we require for any $\mathcal{U} = \{k_1, k_2, \cdots, k_n\} \subset [K]$
\begin{eqnarray}
\left[\begin{array}{c}
    {\bf{V}}_{k_1}^{n\leftarrow m} {\bf{H}}_{k_1}^{m}\\
    {\bf{V}}_{k_2}^{n\leftarrow m} {\bf{H}}_{k_2}^{m}\\
    \vdots\\
    {\bf{V}}_{k_n}^{n\leftarrow m} {\bf{H}}_{k_n}^{m}
\end{array}
\right]_{K!\times K!} \triangleq {\bf F}_{\mathcal{U}}^m, \forall m \in [n-1]
~\mbox{and}~
\left[\begin{array}{c}
    {\bf{H}}^n_{k_1}\\
    {\bf{H}}^n_{k_2}\\
    \vdots\\
    {\bf{H}}^n_{k_n}
\end{array}
\right]_{K!\times K!} \triangleq {\bf H}^n_{\mathcal{U}}
~\mbox{have full rank}
\label{eq:ach_hg}
\end{eqnarray}
so that
\begin{eqnarray}
H \left(\left(Z_k^n\right)_{k \in \mathcal{U}}\right) 
&\overset{(\ref{eq:ach_zkng})}{=}& 
H\left( \left( {\bf{H}}_k^{n} S^n\right)_{k \in \mathcal{U}} \right)  
+ \sum_{m\in[n-1]} H \left( \left( {\bf{V}}_k^{n \leftarrow m}{\bf{H}}_k^{m} S^m\right)_{k \in \mathcal{U}} \right) \\
&=& \left( \mbox{rank}({\bf{H}}_\mathcal{U}^{n}) + \sum_{m \in [n-1]} \mbox{rank}({\bf{F}}_\mathcal{U}^{m}) \right) \log_2 q \\
&\overset{(\ref{eq:ach_hg})}{=}& n K! \log_2 q = n L.
\label{eq:ach_zg}
\end{eqnarray}
Thus $(\ref{generic})$ is guaranteed when $|\mathcal{U}| = n$. The cases where $\mathcal{U} \neq n$ follow in a straightforward manner (see the explanation in Footnote $1$).

{\it Step 4:} Finally, we show that there exist matrices $\left( {\bf H}_k^n \right)_{k,n\in[K]}, \left( {\bf V}_k^{n_2 \leftarrow n_1} \right)_{k,n_1,n_2\in[K], n_2 > n_1}$ that satisfy (\ref{eq:ach_hg}).
Draw each element of the matrices ${\bf H}_k^n, {\bf V}_k^{n_2 \leftarrow n_1}$ independently and uniformly from $\mathbb{F}_q$. Denote the vector that contains all such elements as $\vec{v}$. View the determinant of each matrix in (\ref{eq:ach_hg}) as a polynomial in $\vec{v}$ and consider the product of all such polynomials, denoted by $f(\vec{v})$.
$f(\vec{v})$ is product of $\sum_{n\in[K]} n \binom{K}{n}$ polynomials, each of which has degree at most $K!$, so the degree of $f(\vec{v})$ is at most $K! \sum_{n\in[K]} n \binom{K}{n}$.

$f(\vec{v})$ is not the zero polynomial (whose proof is straightforward as we may find realizations of ${\bf H}_k^n, {\bf V}_k^{n_2 \leftarrow n_1}$ such that each matrix in (\ref{eq:ach_hg}) is the identity matrix following the proof of Lemma \ref{lemma:full}), so we can apply the Schwartz–Zippel lemma to obtain
$\Pr(f(\vec{v}) = 0) < 1.$
Therefore, there exists at least one assignment of ${\bf H}_k^n, {\bf V}_k^{n_2 \leftarrow n_1}$ so that all matrices in (\ref{eq:ach_hg}) have full rank and thus the generated variables are indeed MDS.

\section{Proof of Theorem \ref{theorem2}: Converse}\label{sec:sum_converse}
To illustrate the ideas in a simpler setting, we first consider the setting where $K=4$ and then generalize its proof to arbitrary $K$.

\subsection{Example: $K = 4$ and $R_Z \geq 1 + 1/2 + 1/3$
}
Similar to the converse proof of Theorem \ref{theorem1}, the proof here is also recursive. However, the recursion is significantly more challenging technically (on the mutual information terms) and connecting the mutual information terms to the key rate is also much less obvious.

{\it Step 1}: Consider $I(Z_2;Z_1)$ and we will show that $I(Z_2;Z_1)\geq L\log_2 q$. The proof of this inequality will involve two selected users, i.e., $\mathcal{U} = \{1,2\}$. We first relate the mutual information term on key variables to an entropy term on the inputs and messages.
\begin{eqnarray}
I\left(Z_2;Z_1\right) &\overset{(\ref{input_ind})}{=}& I\left(W_2,Z_2;W_1,Z_1\right)\label{ex2_con_step1_eq3}\\
&\overset{(\ref{message})}{\geq}& I\left(W_2,Z_2;W_1,X_1^{\{1,2\}}\right)\label{ex2_con_step1_eq4}\\
&\geq& I\left(W_2,Z_2;W_1\Big|X_1^{\{1,2\}}\right)\\
&=& H\left(W_1\Big| X_1^{\{1,2\}}\right) - 
H\left(W_1\Big| X_1^{\{1,2\}}, W_2,Z_2\right)
\label{ex2_con_step1_eq5}\\
&\overset{(\ref{corr})}{=}& H\left(W_1\Big| X_1^{\{1,2\}}\right)\label{ex2_con_step1b}
\end{eqnarray}
where (\ref{ex2_con_step1_eq3}) is due to the independence of $(W_k)_{k\in [K]}$ and $(Z_k)_{k \in [K]}$ (see Lemma \ref{lemma_identity2} for a detailed proof), 
(\ref{ex2_con_step1_eq4}) follows from the fact that $X_1^{\{1,2\}}$ is determined by $W_1,Z_1$, and the second term of (\ref{ex2_con_step1_eq5}) is 0 because 1) $X_{2}^{\{1,2\}}$ can be obtained from $W_2, Z_2$, 2) $W_1+W_2$ can be decoded from $X_{1}^{\{1,2\}},X_{2}^{\{1,2\}}$, and 3) from $W_1+W_2$ and $W_2$, $W_1$ can be recovered. 

Next, we show that $H\left(W_1 \Big| X_1^{\{1,2\}}\right)\geq L\log_2 q$ so the desired inequality is obtained. The intuition of this inequality is obvious, i.e., from the security constraint (\ref{sec}), $X_1^{\{1,2\}}$ should not reveal anything about $W_1$.
\begin{eqnarray}
H \left(W_1 \Big|X_1^{\{1,2\}}\right) 
&=& H \left(W_1\right) - H \left(W_1\right) + H \left(W_1 \Big|X_1^{\{1,2\}}\right)\\
&\overset{(\ref{input_ind})(\ref{input_size_L})}{\geq}& L\log_2 q - H \left(W_1 | W_1+W_2\right) + H \left(W_1 \Big| X_1^{\{1,2\}},X_2^{\{1,2\}}, W_1+W_2\right)\label{ex2_con_step1_eq1}\\
&=& L\log_2 q - I \left(W_1 ; X_1^{\{1,2\}},X_2^{\{1,2\}} \Big| W_1+W_2\right)\\
&\geq& L \log_2 q - \underbrace{I \left(W_1,W_2; X_1^{\{1,2\}},X_2^{\{1,2\}} \Big| W_1+W_2\right)}_{\overset{(\ref{sec})}{=}0} ~=~ L\log_2 q \label{ex2_con_step1}
\end{eqnarray}
where the second term of (\ref{ex2_con_step1_eq1}) follows from the fact that $W_1,W_2$ are independent and uniform. 

Summarizing what we have proved and by symmetry, we have
\begin{eqnarray}
I\left(Z_i;Z_j\right) \geq L\log_2 q, \forall i,j \in [K], i\neq j. 
\label{ex2_con_step1a}
\end{eqnarray}

{\it Step 2}: Consider $I(Z_2;Z_1,Z_3)+I(Z_3;Z_1,Z_2)$, which will be proved to be no smaller than $3L\log_2 q$. 
Note that $I(Z_2;Z_1,Z_3)+I(Z_3;Z_1,Z_2) = I(Z_2;Z_3) + [I(Z_2;Z_1|Z_3)+I(Z_3;Z_1,Z_2)]$, where the first term has been proved to be no smaller than $L\log_2 q$ in {\it Step 1}, so we are left to prove $I\left(Z_3;Z_1,Z_2\right)+I\left(Z_2;Z_1|Z_3\right) \geq 2L\log_2 q$. To this end, we select three users, i.e., $\mathcal{U} = \{1,2, 3\}$. Similar to the previous step, we first relate the target mutual information sum to a sum of entropy terms on the inputs and messages.
\begin{eqnarray}
&& I\left(Z_3;Z_1,Z_2\right)+I\left(Z_2;Z_1|Z_3\right) \notag\\
&\overset{(\ref{input_ind})}{=}& I\left(W_3,Z_3;W_1,Z_1,W_2,Z_2\right) +  I\left(W_2,Z_2;W_1,Z_1|W_3,Z_3\right)\\
&\overset{(\ref{message})}{\geq}& I\left(W_3,Z_3;W_1,X_1^{\{1,2,3\}},W_2,X_2^{\{1,2,3\}}\right) +  I\left(W_2,X_2^{\{1,2,3\}};W_1,X_1^{\{1,2,3\}} \Big|W_3,Z_3\right)\\
&\geq& I\left(W_3,Z_3;W_1,X_1^{\{1,2,3\}}\right) + I\left(W_3,Z_3;W_2,X_2^{\{1,2,3\}} \Big|W_1,X_1^{\{1,2,3\}}\right)\notag\\
&&~+ I\left(W_2,X_2^{\{1,2,3\}};W_1\Big|W_3,Z_3,X_1^{\{1,2,3\}}\right)\\
&\overset{(\ref{message})}{\geq}&
I\left(W_3,Z_3;W_1\Big|X_1^{\{1,2,3\}}\right) + I\left(W_3,X_3^{\{1,2,3\}};W_2\Big|W_1,X_1^{\{1,2,3\}},X_2^{\{1,2,3\}}\right) \notag\\
&&
~+H\left(W_1\Big|W_3,Z_3,X_1^{\{1,2,3\}}\right)-\underbrace{H\left(W_1\Big|W_3,Z_3,X_1^{\{1,2,3\}},W_2,X_2^{\{1,2,3\}}\right)}_{\overset{(\ref{message})(\ref{corr})}{=}0}\\
&=&
H\left(W_1\Big|X_1^{\{1,2,3\}}\right) - H\left(W_1\Big|X_1^{\{1,2,3\}},W_3,Z_3\right) +
H\left(W_2\Big|W_1,X_1^{\{1,2,3\}},X_2^{\{1,2,3\}}\right)\notag\\
&&~-
\underbrace{H\left(W_2\Big|W_1,X_1^{\{1,2,3\}},X_2^{\{1,2,3\}},W_3,X_3^{\{1,2,3\}}\right)}_{\overset{(\ref{corr})}{=}0}+~H\left(W_1\Big|W_3,Z_3,X_1^{\{1,2,3\}}\right) \label{eq:common}\\
&=& H\left(W_1\Big|X_1^{\{1,2,3\}}\right)+H\left(W_2\Big|W_1,X_1^{\{1,2,3\}},X_2^{\{1,2,3\}}\right). \label{ex2_con_step2b}
\end{eqnarray}

Next, following the proof of (\ref{ex2_con_step1}), we may show using the security constraint that (details are deferred to Lemma \ref{lemma3})
\begin{eqnarray}
&&H\left(W_1 \Big|X_1^{\{1,2,3\}}\right)\geq L\log_2 q, \notag \label{ex2_con_step2_eq1}\\
&&H\left(W_2 \Big|W_1,X_1^{\{1,2,3\}},X_2^{\{1,2,3\}}\right) \geq L\log_2 q.\label{ex2_con_step2_eq2}
\end{eqnarray}

Combining (\ref{ex2_con_step2b}), (\ref{ex2_con_step2_eq2}) and by symmetry, we have for all distinct $i_1, i_2, i_3 \in [K]$ 
\begin{eqnarray}
I\left(Z_{i_2};Z_{i_1},Z_{i_3}\right)+I\left(Z_{i_3};Z_{i_1}, Z_{i_2}\right) 
&=& \big[ \underbrace{I\left(Z_{i_3};Z_{i_1},Z_{i_2}\right)+I\left(Z_{i_2};Z_{i_1}|Z_{i_3}\right)}_{\overset{(\ref{ex2_con_step2b}), (\ref{ex2_con_step2_eq2})}{\geq} L\log_2 q+L\log_2 q} \big] + \underbrace{I\left(Z_{i_2};Z_{i_3}\right)}_{\overset{(\ref{ex2_con_step1a})}{\geq} L\log_2 q} \\
&\geq& 3L\log_2 q.
\label{ex2_con_step2}
\end{eqnarray}

\begin{remark}
Crucially for the above proof, we need to consider the inequality on the sum of (conditional) mutual information terms, $I\left(Z_3;Z_1,Z_2\right)+I\left(Z_2;Z_1|Z_3\right) \geq 2L\log_2 q$ while no meaningful and useful converse bounds can be proved for each individual term. The intuition is to capture the total correlation of all keys used in selecting three users, which could appear in either $I\left(Z_3;Z_1,Z_2\right)$ or $I\left(Z_2;Z_1|Z_3\right)$. The critical step in the above derivation is 
to extract a common term $H\left(W_1\Big|W_3,Z_3,X_1^{\{1,2,3\}}\right)$ that cancels (see (\ref{eq:common})).
\end{remark}

{\it Step 3.1}: Consider $I\left(Z_4;Z_1,Z_2,Z_3\right) + I\left(Z_3;Z_1,Z_2|Z_4\right) + 
I\left(Z_2;Z_1|Z_3,Z_4\right) \geq 3L\log_2 q$, which is an important intermediate step to proceed to the recursion on key rate (refer to {\it Step 3.2} and (\ref{eq:ex_rec})). Set $\mathcal{U} = \{1,2, 3, 4\}$. Similar to the previous steps, we first relate the target mutual information sum to a sum of entropy terms on the inputs and messages.

Following the proof of (\ref{ex2_con_step2b}), we may prove
\begin{eqnarray}
&& I\left(Z_3;Z_1,Z_2|Z_4\right) + I\left(Z_2;Z_1|Z_3,Z_4\right)\notag\\
&\geq&
H\left(W_1 \Big|W_4,Z_4,X_1^{\{1,2,3,4\}}\right) + H\left(W_2 \Big|W_4,Z_4,W_1,X_1^{\{1,2,3,4\}},X_2^{\{1,2,3,4\}}\right) \label{eq:ex23}
\end{eqnarray}
and 
\begin{eqnarray}
&& I\left(Z_4;Z_1,Z_2,Z_3\right)\notag\\
&\geq& H\left(W_1 \Big| X_1^{\{1,2,3,4\}}\right)
+H\left(W_2 \Big| W_1,X_1^{\{1,2,3,4\}},X_2^{\{1,2,3,4\}}\right)\notag\\
&&
~+ H\left(W_3 \Big| W_2,X_2^{\{1,2,3,4\}},W_1,X_1^{\{1,2,3,4\}},X_3^{\{1,2,3,4\}}\right)\notag\\
&&
~-H\left(W_1 \Big| X_1^{\{1,2,3,4\}},W_4,Z_4\right)-H\left(W_2 \Big| W_1,X_1^{\{1,2,3,4\}},X_2^{\{1,2,3,4\}},W_4,Z_4\right) \label{ex2_con_step3_eq3}
\end{eqnarray}
where the detailed proof is deferred to Lemma \ref{lemma:mk}. Adding (\ref{eq:ex23}) and (\ref{ex2_con_step3_eq3}) (to cancel the deliberately split terms), we have
\begin{eqnarray}
&&I\left(Z_4;Z_1,Z_2,Z_3\right) + I\left(Z_3;Z_1,Z_2|Z_4\right) + 
I\left(Z_2;Z_1|Z_3,Z_4\right)\notag\\
&\geq&
H\left(W_1 \Big| X_1^{\{1,2,3,4\}}\right)
+H\left(W_2 \Big| W_1,X_1^{\{1,2,3,4\}},X_2^{\{1,2,3,4\}}\right)\notag\\
&&~+ H\left(W_3 \Big| W_2,X_2^{\{1,2,3,4\}},W_1,X_1^{\{1,2,3,4\}},X_3^{\{1,2,3,4\}}\right)\label{ex2_con_step3b}\\
&\geq& 3L \log_2 q \label{ex2_con_step3a}
\end{eqnarray}
where (\ref{ex2_con_step3a}) follows by proving each term of (\ref{ex2_con_step3b}) is no smaller than $L\log_2 q$ (the proof is deferred to Lemma \ref{lemma3}). Note that by symmetry, (\ref{ex2_con_step3a}) holds for any $4$ distinct $Z_k$ variables.

{\it Step 3.2:} Finally, we are ready to connect the mutual information inequalities established in the previous steps to the key rate.
Note that the set of all permutations of $\{1,2,3,4\}$ is denoted as $\mathcal{S}_4 = (\pi_n)_{n \in [4!]}$. 
\begin{eqnarray}
&& 4!\times 3 L_Z\log_2 q \overset{(\ref{h1})}{\geq} \sum_{\pi \in \mathcal{S}_4} 3H\left(Z_{\pi(1)}\right) \geq \sum_{\pi \in \mathcal{S}_4} 3 I\left(Z_{\pi(1)};Z_{\pi(2)},Z_{\pi(3)},Z_{\pi(4)}\right) \label{eq:ex_rec}\\
&=& \sum_{\pi \in \mathcal{S}_4} \left[ I\left(Z_{\pi(1)};Z_{\pi(2)},Z_{\pi(3)},Z_{\pi(4)}\right) + I\left(Z_{\pi(1)};Z_{\pi(3)},Z_{\pi(4)} \big|Z_{\pi(2)}\right) + I\left(Z_{\pi(1)};Z_{\pi(4)} \big|Z_{\pi(2)},Z_{\pi(3)}\right)\right]\notag\\
&&~+\sum_{\pi \in \mathcal{S}_4}I\left(Z_{\pi(1)};Z_{\pi(2)},Z_{\pi(3)}\right)+\sum_{\pi \in \mathcal{S}_4}I\left(Z_{\pi(1)};Z_{\pi(2)}\right)\label{ex2_con_step3_eq4}\\
&=& \sum_{\pi \in \mathcal{S}_4} \underbrace{\left[ I\left(Z_{\pi(4)};Z_{\pi(1)},Z_{\pi(2)},Z_{\pi(3)}\right) + I\left(Z_{\pi(3)};Z_{\pi(1)},Z_{\pi(2)} \big|Z_{\pi(4)}\right) + I\left(Z_{\pi(2)};Z_{\pi(1)} \big|Z_{\pi(3)},Z_{\pi(4)}\right)\right]}_{\overset{(\ref{ex2_con_step3a})}{\geq}3L\log_2 q}\notag\\
&&~+ \frac{1}{2}\sum_{\pi \in \mathcal{S}_4} \underbrace{\left[I\left(Z_{\pi(2)};Z_{\pi(1)},Z_{\pi(3)}\right)+I\left(Z_{\pi(3)};Z_{\pi(1)},Z_{\pi(2)}\right)\right]}_{\overset{(\ref{ex2_con_step2})}{\geq}3L\log_2 q}+ \sum_{\pi \in \mathcal{S}_4} \underbrace{I\left(Z_{\pi(1)};Z_{\pi(2)}\right)}_{\overset{(\ref{ex2_con_step1a})}{\geq}L\log_2 q}\label{ex2_con_step3_eq5}\\
&\geq& 4! \times 3L\log_2 q + 4! \times 3/2 \times L\log_2 q + 4! \times L\log_2 q\label{ex2_con_step3_eq6}\\
&\Rightarrow& R_Z = L_Z/L \geq 1+1/2+1/3
\end{eqnarray}
where in (\ref{ex2_con_step3_eq4}), we split the same mutual information term using three different ways to apply the three mutual information inequalities proved in {\it Step 1}, {\it Step 2}, and {\it Step 3.1}. 
The change of indices in (\ref{ex2_con_step3_eq5}) follows from the fact that we are considering all permutations.

\begin{remark}
While the target is the key rate, captured by $H(Z_1)$, our proof is based on relating it to the mutual information $I(Z_1; Z_2,\cdots, Z_K)$, and expanding the mutual information terms and combining the split terms in a highly non-trivial manner that requires a technical inequality for $K$ selected users (see the first term of (\ref{ex2_con_step3_eq5}) and Lemma \ref{lemma:mk}) and a recursive use of bounds on $I(Z_1; Z_2,\cdots, Z_m)$ for all $m \in \{2,\cdots,K-1\}$ (see the second to last terms of (\ref{ex2_con_step3_eq5}) and Lemma \ref{lemma:mu}).
\end{remark}

\subsection{General Proof: $R_Z \geq 1 + 1/2 + \cdots 1/(K-1)$}
Let us start with a useful identity that allows us to include input variables to key variables.

\begin{lemma}\label{lemma_identity2}
For any disjoint $\mathcal{U}_1,\mathcal{U}_2,\mathcal{U}_3  \subset \mathcal{K}$, we have
\begin{eqnarray}
I\left(\left(Z_i\right)_{i \in \mathcal{U}_1};\left(Z_i\right)_{i \in \mathcal{U}_2} \big| \left(Z_i\right)_{i \in \mathcal{U}_3}\right)
= I\left(\left(W_i, Z_i\right)_{i \in \mathcal{U}_1}; \left(W_i, Z_i\right)_{i \in \mathcal{U}_2} \big| \left(W_i, Z_i\right)_{i \in \mathcal{U}_3}\right). \label{lemma_identity2_eq}
\end{eqnarray}
\end{lemma}
{\it Proof:} \begin{eqnarray}
&& I\left(\left(W_i, Z_i\right)_{i \in \mathcal{U}_1}; \left(W_i, Z_i\right)_{i \in \mathcal{U}_2} \big| \left(W_i, Z_i\right)_{i \in \mathcal{U}_3}\right) \notag\\
&=& H\left(\left(W_i, Z_i\right)_{i \in \mathcal{U}_1}, \left(W_i, Z_i\right)_{i \in \mathcal{U}_3}\right) + H\left(\left(W_i, Z_i\right)_{i \in \mathcal{U}_2}, \left(W_i, Z_i\right)_{i \in \mathcal{U}_3}\right) \notag\\
&& ~- H\left(\left(W_i, Z_i\right)_{i \in \mathcal{U}_1}, \left(W_i, Z_i\right)_{i \in \mathcal{U}_2},\left(W_i, Z_i\right)_{i \in \mathcal{U}_3}\right) - H\left( \left(W_i, Z_i\right)_{i \in \mathcal{U}_3}\right) \\
&\overset{(\ref{input_ind})}{=}& H\left(\left(Z_i\right)_{i \in \mathcal{U}_1}, \left(Z_i\right)_{i \in \mathcal{U}_3}\right) + H\left(\left(Z_i\right)_{i \in \mathcal{U}_2}, \left(Z_i\right)_{i \in \mathcal{U}_3}\right) \notag\\
&& ~- H\left(\left(Z_i\right)_{i \in \mathcal{U}_1}, \left(Z_i\right)_{i \in \mathcal{U}_2},\left(Z_i\right)_{i \in \mathcal{U}_3}\right) - H\left( \left(Z_i\right)_{i \in \mathcal{U}_3}\right) \label{eq:id2}\\
&=& I\left(\left(Z_i\right)_{i \in \mathcal{U}_1};\left(Z_i\right)_{i \in \mathcal{U}_2} \big| \left(Z_i\right)_{i \in \mathcal{U}_3}\right)
\end{eqnarray}
where the $W_k$ terms are fully cancelled in (\ref{eq:id2}) because $W_k$ is independent of $Z_k$ and $W_k$ is independent for disjoint sets of indicies $k$.

\hfill\QED

Next, we state two properties that will be used to provide bounds on mutual information terms. The first property builds upon the security constraint and states that certain conditional entropy term is no smaller than $L\log_2 q$.

\begin{lemma}\label{lemma3} 
For any set $\mathcal{U} \subset [K]$, we have
\begin{eqnarray}\label{lemma3_eq}
H \left(W_k \big|X_k^{\mathcal{U}},(W_i,X_i^{\mathcal{U}})_{i \in \mathcal{V}})\right) \geq L\log_2 q, ~\forall k \in \mathcal{U}, \forall \mathcal{V} \subsetneq \mathcal{U} \backslash \{k\}. 
\end{eqnarray}
\end{lemma}
{\it Proof:
}
\begin{eqnarray}
&& H \left(W_k \big|X_k^{\mathcal{U}},(W_i,X_i^{\mathcal{U}})_{i \in \mathcal{V}})\right) \notag \\
&=& H \left(W_k\right) - H \left(W_k\right) + H \left(W_k \big|X_k^{\mathcal{U}},(W_i,X_i^{\mathcal{U}})_{i \in \mathcal{V}}\right)\\
&\overset{(\ref{input_ind})(\ref{input_size_L})}{\geq}& L\log_2 q - H \left(W_k\Bigg|\sum_{i \in \mathcal{U}\backslash \mathcal{V}} W_i, (W_i)_{i \in \mathcal{V}}\right) + H \left(W_k \Bigg| \left(X_i^{\mathcal{U}}\right)_{i \in \mathcal{U}},\left(W_i\right)_{i \in \mathcal{V}},\sum_{i \in \mathcal{U}\backslash \mathcal{V}} W_i\right)\label{lemma3_eq1}\\
&=& L\log_2 q - I \left(W_k ; \left(X_i^{\mathcal{U}}\right)_{i \in \mathcal{U}} \Bigg| \left(W_i\right)_{i \in \mathcal{V}},\sum_{i \in \mathcal{U}\backslash \mathcal{V}} W_i\right)\\
&=& L\log_2 q - I \left(W_k ; \left(X_i^{\mathcal{U}}\right)_{i \in \mathcal{U}} \Bigg| \left(W_i\right)_{i \in \mathcal{V}},\sum_{i \in \mathcal{U}} W_i\right)\\
&\geq& L\log_2 q - \underbrace{I \left(\left(W_i\right)_{i \in \mathcal{U}}; \left(X_i^{\mathcal{U}}\right)_{i \in \mathcal{U}} \Bigg| \sum_{i \in \mathcal{U}} W_i\right)}_{\overset{(\ref{sec})}{=}0} ~=~ L\log_2 q \label{eq:temp2}
\end{eqnarray}
where (\ref{lemma3_eq1}) uses the independence and uniformity of $(W_k)_{k \in \mathcal{U}}$ and the fact that $\mathcal{V} \cup \{k\} \subsetneq \mathcal{U}$; (\ref{eq:temp2}) uses the fact that $\mathcal{V} \cup \{k\} \subset \mathcal{U}$. 

\hfill\QED

The second property relates a mutual information sum to the sum of conditional entropy terms considered in the previous lemma, combining with which gives us the desired bound on the mutual information sum. We use the notation $[i: j] = \{i, i+1, \cdots, j\}$ if $i \leq j$ and otherwise $[i:j]$ is an empty set. Recall that $[i]$ denotes the set $\{1,\cdots, i\}$.

\begin{lemma}\label{lemma:mk}
For any set $\mathcal{U} = \{k_1,k_2,\cdots,k_{|\mathcal{U}|}\} \subset [K]$, $|\mathcal{U}|\geq 2$, we have
\begin{eqnarray}
 &&\sum_{i \in [2:m]} I\left(Z_{k_i}; \left(Z_{k_j}\right)_{j \in [i-1]} \Big|\left(Z_{k_j}\right)_{j \in [i+1:|\mathcal{U}|]}\right)\notag\\
 &\geq& \sum_{i \in [m-1]} H \left(W_{k_i} \Big| (W_{k_j},X_{k_j}^{\mathcal{U}})_{j \in [i-1]}, X_{k_i}^{\mathcal{U}},\left(W_{k_j},Z_{k_j}\right)_{j \in [m+1:|\mathcal{U}|]}\right), \forall m \in [2:|\mathcal{U}|]. \label{lemma4_stat_eq} \\
&& \sum_{i \in [2:|\mathcal{U}|]} I\left(Z_{k_i}; \left(Z_{k_j}\right)_{j \in [i-1]} \Big|\left(Z_{k_j}\right)_{j \in [i+1:|\mathcal{U}|]}\right) 
\geq
(|\mathcal{U}|-1)L\log_2 q.
\label{lemma4_eq1}
\end{eqnarray}
\end{lemma}
{\it Proof:} First, consider (\ref{lemma4_stat_eq}), whose proof is by induction on $m$.

{\it Base case:} We show that (\ref{lemma4_stat_eq}) holds when $m=2$. 
\begin{eqnarray}
&& I\left(Z_{k_2};Z_{k_1} \Big| \left(Z_{k_j}\right)_{j \in [3:|\mathcal{U}|]}\right)  \notag\\
&\overset{(\ref{lemma_identity2_eq})}{=}& I\left(W_{k_2},Z_{k_2};W_{k_1},Z_{k_1} \Big|\left(W_{k_j},Z_{k_j}\right)_{j \in [3:|\mathcal{U}|]}\right) \\
&\overset{(\ref{message})}{\geq}& I\left(W_{k_2},Z_{k_2};W_{k_1},X_{k_1}^{\mathcal{U}} \Big|\left(W_{k_j},Z_{k_j}\right)_{{j} \in [3:|\mathcal{U}|]}\right)\\
&\geq& I\left(W_{k_2},Z_{k_2};W_{k_1} \Big|\left(W_{k_j},Z_{k_j}\right)_{j \in [3:|\mathcal{U}|]},X_{k_1}^{\mathcal{U}}\right)\\
&=& H \left(W_{k_1} \Big|\left(W_{k_j},Z_{k_j}\right)_{j \in [3:|\mathcal{U}|]},X_{k_1}^{\mathcal{U}}\right) - \underbrace{H\left(W_{k_1} \Big|\left(W_{k_j},Z_{k_j}\right)_{j \in [3:|\mathcal{U}|]},X_{k_1}^{\mathcal{U}},W_{k_2},Z_{k_2}\right)}_{\overset{(\ref{message})(\ref{corr})}{=}0}\\
&=& H \left(W_{k_1} \Big| X_{k_1}^{\mathcal{U}}, \left(W_{k_j},Z_{k_j}\right)_{j \in [3:|\mathcal{U}|]}\right).
\end{eqnarray}

{\it Induction step:} Suppose (\ref{lemma4_stat_eq}) holds for $m = M, 2\leq M \leq |\mathcal{U}|-1$, then we show that  (\ref{lemma4_stat_eq}) also holds for $m = M+1$. When $m = M+1$, LHS of (\ref{lemma4_stat_eq}) contains one more term when compared to that when $m = M$. Consider this additional term.
\begingroup
\allowdisplaybreaks
\begin{eqnarray}
&& I\left(Z_{k_{M+1}};\left(Z_{k_j}\right)_{j \in [M]} \Big| \left(Z_{k_j}\right)_{j \in [M+2:|\mathcal{U}|]}\right) \label{lemma4_eq2}\\
&\overset{(\ref{lemma_identity2_eq})}{=}& I\left(W_{k_{M+1}},Z_{k_{M+1}};\left(W_{k_j},Z_{k_j}\right)_{j \in [M]} \Big| \left(W_{k_j},Z_{k_j}\right)_{j \in [M+2:|\mathcal{U}|]}\right) \label{lemma4_eq3}\\
&\overset{(\ref{message})}{\geq}& I\left(W_{k_{M+1}},Z_{k_{M+1}};\left(W_{k_j},X_{k_j}^{\mathcal{U}}\right)_{j \in [M]} \bigg| \left(W_{k_j},Z_{k_j}\right)_{j \in [M+2:|\mathcal{U}|]}\right) \label{lemma4_eq4}\\
&=&
\sum_{i \in [M]} I\left(W_{k_{M+1}},Z_{k_{M+1}};W_{k_i},X_{k_i}^{\mathcal{U}} \Big|  \left(W_{k_j},X_{k_j}^{\mathcal{U}}\right)_{j \in [i-1]}, \left(W_{k_j},Z_{k_j}\right)_{j \in [M+2:|\mathcal{U}|]}\right) \\
&\geq&
\sum_{i \in [M]} I\left(W_{k_{M+1}},Z_{k_{M+1}};W_{k_i} \bigg|  \left(W_{k_j},X_{k_j}^{\mathcal{U}}\right)_{j \in [i-1]},X_{k_i}^{\mathcal{U}},\left(W_{k_j},Z_{k_j}\right)_{j \in [M+2:|\mathcal{U}|]}\right) \\
&=&  \sum_{i \in [M]} H\left(W_{k_i} \bigg|
 \left(W_{k_j},X_{k_j}^{\mathcal{U}}\right)_{j \in [i-1]},X_{k_i}^{\mathcal{U}},\left(W_{k_j},Z_{k_j}\right)_{j \in [M+2:|\mathcal{U}|]}\right) \notag\\
&&~- \sum_{i \in [M]} H\left(W_{k_i} \bigg| \left(W_{k_j},X_{k_j}^{\mathcal{U}}\right)_{j \in [i-1]},X_{k_i}^{\mathcal{U}}, \left(W_{k_j},Z_{k_j}\right)_{j \in [M+1:|\mathcal{U}|]}\right) \\
&=&  
\sum_{i \in [M]} H\left(W_{k_i} \bigg|  \left(W_{k_j},X_{k_j}^{\mathcal{U}}\right)_{j \in [i-1]},X_{k_i}^{\mathcal{U}},\left(W_{k_j},Z_{k_j}\right)_{j \in [M+2:|\mathcal{U}|]}\right) \notag\\
&&~- 
\underbrace{H\left(W_{k_M} \bigg|\left(W_{k_j},X_{k_j}^{\mathcal{U}}\right)_{j \in [M-1]},X_{k_M}^{\mathcal{U}}, \left(W_{k_j},Z_{k_j}\right)_{j \in [M+1:|\mathcal{U}|]}\right)}_{\overset{(\ref{message})(\ref{corr})}{=}0} \notag\\
&&~- \underbrace{ \sum_{i \in [M-1]} H\left(W_{k_i} \bigg| \left(W_{k_j},X_{k_j}^{\mathcal{U}}\right)_{j \in [i-1]},X_{k_i}^{\mathcal{U}}, \left(W_{k_j},Z_{k_j}\right)_{j \in [M+1:|\mathcal{U}|]}\right) }_{\leq \sum_{i \in [2:M]} I\left(Z_{k_i}; \left(Z_{k_j}\right)_{j \in [i-1]} \Big|\left(Z_{k_j}\right)_{j \in [i+1:|\mathcal{U}|]}\right) ~\mbox{\footnotesize (induction assumption)}} \\
&\Rightarrow&  \sum_{i \in [2:M+1]} I\left(Z_{k_i}; \left(Z_{k_j}\right)_{j \in [i-1]} \Big|\left(Z_{k_j}\right)_{j \in [i+1:|\mathcal{U}|]}\right) \notag\\
&\geq& \sum_{i \in [M]} H\left(W_{k_i} \bigg|  \left(W_{k_j},X_{k_j}^{\mathcal{U}}\right)_{j \in [i-1]},X_{k_i}^{\mathcal{U}},\left(W_{k_j},Z_{k_j}\right)_{j \in [M+2:|\mathcal{U}|]}\right)
\end{eqnarray}
so that we arrive at (\ref{lemma4_stat_eq}) when $n = M+1$ and the proof of (\ref{lemma4_stat_eq}) by induction is complete.

Second, consider (\ref{lemma4_eq1}), which follows directly from Lemma \ref{lemma3} and (\ref{lemma4_stat_eq}). Set $m=|\mathcal{U}|$ in (\ref{lemma4_stat_eq}), then
\begin{eqnarray}
&& \sum_{i \in [2:|\mathcal{U}|]} I\left(Z_{k_i}; \left(Z_{k_j}\right)_{j \in [i-1]} \Big|\left(Z_{k_j}\right)_{j \in [i+1:|\mathcal{U}|]}\right) \notag\\
&\overset{(\ref{lemma4_stat_eq})}{\geq}& \sum_{i \in [|\mathcal{U}|-1]} H \left(W_{k_i} \Big| (W_{k_j},X_{k_j}^{\mathcal{U}})_{j \in [i-1]}, X_{k_i}^{\mathcal{U}}\right)\label{lemma4_eq0}\\
&\overset{(\ref{lemma3_eq})}{\geq}& 
(|\mathcal{U}|-1)L\log_2 q.
\end{eqnarray}
\hfill\QED

We are now ready to recursively bound the mutual information (correlation) between one key and any number of other keys, in the following lemma.

\begin{lemma}\label{lemma:mu}
For any set $\mathcal{U} = \{k_1,k_2,\cdots,$ $k_{|\mathcal{U}|}\} \subset [K]$, $|\mathcal{U}|\geq 2$, we have
\begin{eqnarray}
\frac{1}{|\mathcal{U}|!} \sum_{\pi \in \mathcal{S}_{|\mathcal{U}|}} I \left(Z_{k_{\pi (1)}}; \left( Z_{k_{\pi (j)}}\right)_{j \in [2: |\mathcal{U}|]}\right) \geq \left(1 + \frac{1}{2} +\cdots + \frac{1}{|\mathcal{U}
|-1}\right)L\log_2 q \label{lemma_converse2_eq}
\end{eqnarray}
where $\mathcal{S}_{m}$ is the set of all permutations of $[m]$.
\end{lemma}
{\it Proof:} The proof is based on mathematical induction on $|\mathcal{U}|$.

{\it Base case:} We show that (\ref{lemma_converse2_eq}) holds when $|\mathcal{U}|=2$. Consider any $\mathcal{U}\subset [K]$ where $|\mathcal{U}|=2$.
\begin{eqnarray}
\frac{1}{2!} \sum_{\pi \in \mathcal{S}_{2}} I \left(Z_{k_{\pi (1)}};  Z_{k_{\pi (2)}} \right) &=&
\frac{1}{2} \left[I \left(Z_{k_1};Z_{k_2}\right) + I \left(Z_{k_2};Z_{k_1}\right)\right] \\
&\overset{(\ref{lemma4_eq1})}{\geq}& 
L\log_2 q.
\end{eqnarray}

{\it Induction  step:} Suppose (\ref{lemma_converse2_eq}) holds for 
$|\mathcal{U}|\in [2:M], 2\leq M \leq K-1$, then we show that (\ref{lemma_converse2_eq}) also holds for $|\mathcal{U}|= M+1$. Consider any $\mathcal{U}\subset [K]$ where $|\mathcal{U}|=M+1$.
\begingroup
\allowdisplaybreaks
\begin{eqnarray}
&&
\sum_{\pi \in \mathcal{S}_{M+1}}  I \left(Z_{k_{\pi (1)}}; \left( Z_{k_{\pi (j)}}\right)_{j \in [2: M+1]}\right)
= \frac{1}{M} \sum_{\pi \in \mathcal{S}_{M+1}}  M  I \left(Z_{k_{\pi (1)}}; \left( Z_{k_{\pi (j)}}\right)_{j \in [2: M+1]}\right) \label{lemma_converse2_eq1}\\
&=& 
\frac{1}{M} \sum_{\pi \in \mathcal{S}_{M+1}}  \sum_{i \in [2:M+1]}\left[ I \left(Z_{k_{\pi (1)}} ;  \left( Z_{k_{\pi (j)}}\right)_{j \in [2: i-1]}\right)
 \right. \notag\\
&&
~+\left. I \left(Z_{k_{\pi (1)}}; \left( Z_{k_{\pi (j)}}\right)_{j \in [i: M+1]} \bigg|  \left( Z_{k_{\pi (j)}}\right)_{j \in [2: i-1]}\right)
\right]\label{lemma_converse2_eq2}\\
&=& \frac{1}{M}  \sum_{i \in [3:M+1]} \sum_{\pi \in \mathcal{S}_{M+1}}  I \left(Z_{k_{\pi (1)}} ;  \left( Z_{k_{\pi (j)}}\right)_{j \in [2: i-1]}\right)
 \notag\\
&&
~+ \frac{1}{M} \underbrace{\sum_{\pi \in \mathcal{S}_{M+1}}  \sum_{i \in [2:M+1]} I \left(Z_{k_{\pi (1)}}; \left( Z_{k_{\pi (j)}}\right)_{j \in [i: M+1]} \bigg|  \left( Z_{k_{\pi (j)}}\right)_{j \in [2: i-1]}\right)}_{\overset{(\ref{lemma4_eq1})}{\geq}(M+1)!ML\log_2 q} \label{lemma_converse2_eq3}\\
&\geq& \frac{1}{M} \sum_{i \in [3:M+1]} \frac{(M+1)!}{(i-1)!} (i-1)!\left(1 + \frac{1}{2} + \cdots + \frac{1}{i-2}\right)L\log_2 q ~~~~\mbox{(Induction)}
 \notag\\
&& ~+ \frac{1}{M} (M+1)! ML\log_2 q
\label{lemma_converse2_eq4}\\
&=&(M+1)! \left(\frac{1}{M} \sum_{i \in [3:M+1]} \left(1 + \frac{1}{2} + \cdots + \frac{1}{i-2}\right)L + \frac{1}{M}ML \right) \log_2 q
\label{lemma_converse2_eq5}\\
&=&\frac{(M+1)!}{M}\left[M + 1 + \left(1 + \frac{1}{2}\right) +\cdots + \left(1 + \frac{1}{2} + \cdots + \frac{1}{M-1}\right) \right]L \log_2 q \label{lemma_converse2_eq6}\\
&=&(M+1)! \left(1  +\frac{1}{2} + \frac{1}{3} + \cdots + \frac{1}{M}\right)L\log_2 q \label{lemma_converse2_eq7}
\end{eqnarray}
\endgroup
where in (\ref{lemma_converse2_eq2}), we expand the mutual information term in $M$ different ways. Note that when $i=2$, the mutual information term is not expanded and remains, i.e., the first term of (\ref{lemma_converse2_eq2}) does not exist and the second term of (\ref{lemma_converse2_eq2}) has no conditioning.
The first term of (\ref{lemma_converse2_eq4}) follows from the induction assumption that (\ref{lemma_converse2_eq}) holds for $|\mathcal{U}| = i-1$, where $2 \leq |\mathcal{U}| \leq M$. 
The second term of (\ref{lemma_converse2_eq4}) is obtained by using (\ref{lemma4_eq1}) and the fact that we are considering all permutations. To see that (\ref{lemma4_eq1}) can be applied, notice that both (\ref{lemma4_eq1}) and (\ref{lemma_converse2_eq3}) concern the sum of $I(A; B | C)$, where $A$ contains one $Z_k$ term, $B$ contains $u$ $Z_k$ terms, where $1\leq u \leq M$, $C$ contains $M-u$ $Z_k$ terms, and the $Z_k$ terms are all distinct.
The derivation from (\ref{lemma_converse2_eq6}) to (\ref{lemma_converse2_eq7}) is the same as that from (\ref{lemma_converse_eq6}) to (\ref{eq:end}). 

\hfill\QED

The converse proof of $R_Z$ follows directly from Lemma \ref{lemma:mu}. Set $|\mathcal{U}| = K$ in (\ref{lemma_converse2_eq}), i.e., $\mathcal{U} = [K]$, then
\begin{eqnarray}
L_Z\log_2 q &\geq& \frac{1}{K!} \sum_{\pi \in \mathcal{S}_{K}} H\left(Z_{k_{\pi (1)}}\right)\\
&\geq& \frac{1}{K!}
\sum_{\pi \in \mathcal{S}_{K}} I \left(Z_{k_{\pi (1)}}; \left( Z_{k_{\pi (j)}}\right)_{j \in [2: K]}\right) \label{converse2_eq1}\\
&\overset{(\ref{lemma_converse2_eq})}{\geq}& \left(1 + \frac{1}{2} +\cdots + \frac{1}{K-1}\right)L\log_2 q \\
\Rightarrow~ R_Z = \frac{L_Z}{L} &\geq& 1 + \frac{1}{2} +\cdots + \frac{1}{K-1} .
\end{eqnarray}

\section{Proof of Theorem \ref{theorem2}: Achievability}\label{sec:sum_ach}
The achievability proof of secure summation follows immediately from applying MDS variable generation. We first set up what is needed from MDS variable generation and use notations with a tilde symbol. We use $K$-user MDS variable generation for ${\tilde Z}_k^n$ with $n \in [K-1]$. Following (\ref{eq:ach_zkg}), we have
\begin{eqnarray}
{\tilde Z_k} =  \left(\left({\bf{H}}_k^{n}S^n\right)_{n\in[K-1]}\right), \forall k \in [K] \label{eq:sumach_zk}
\end{eqnarray}
where ${\tilde Z}_k$ contains $\sum_{n\in[K-1]}\frac{(K-1)!}{n}$ elements from $\mathbb{F}_{\tilde q}$, ${\tilde q} > K! \sum_{n\in[K]} n \binom{K}{n}$ and following (\ref{eq:ach_zkng}), we may generate
\begin{eqnarray}
~\mbox{$(K,n)$-MDS variables}~{\tilde Z}_k^n \in \mathbb{F}_{\tilde q}^{1 \times (K-1)!}, \forall n \in [K-1], k \in [K] \label{eq:sumach_mds}
\end{eqnarray}
that are linear, so for any $\mathcal{U} \subset [K]$ where $|\mathcal{U}| = n+1$, due to the MDS property (\ref{generic}) there exist full rank matrices ${\bf F}_{u}^{\mathcal{U}} \in \mathbb{F}_{\tilde q}^{(K-1)!\times (K-1)!}$, $u \in \mathcal{U}$ so that
\begin{eqnarray}
\sum_{u \in \mathcal{U}} {\bf F}_{u}^{\mathcal{U}} \left({\tilde Z}_{u}^{n}\right)^T = {\bf 0}_{(K-1)! \times 1} \label{eq:sumach_zu}
\end{eqnarray}
where for any vector $A$, $A^T$ represents its transpose.

We now proceed to consider the secure summation problem, where the field size is fixed to $q$ and $q$ might be smaller than the requirement in MDS variable generation. So we resort to block codes (symbol extensions) and group $B$ input symbols together so that $q^B > K! \sum_{n\in[K]} n \binom{K}{n}$ and set $q^B = {\tilde q}$. View inputs $W_k$ now as elements over the field $\mathbb{F}_{q^B} = \mathbb{F}_{\tilde q}$. Set $L = (K-1)! B$ so that each input $W_k$ consists of $(K-1)!$ symbols over $\mathbb{F}_{\tilde q}$. Set 
\begin{eqnarray}
Z_k = {\tilde Z_k}, \forall k \in [K]
\end{eqnarray}
where ${\tilde Z_k}$ is from (\ref{eq:sumach_zk}). Note that $L_Z=\sum_{n \in [K-1]} L/n$, thus $R_Z = \sum_{n \in [K-1]} 1/n$, as desired. 

For any set of selected users $\mathcal{U} \subset [K]$, where $|\mathcal{U}| = n+1 \geq 2$, set the messages as
\begin{eqnarray}
X_u^{\mathcal{U}} = W_u + {\bf F}_{u}^{\mathcal{U}} \left({\tilde Z}_{u}^{n}\right)^T, \forall u \in \mathcal{U} \label{eq:sumach_xu}
\end{eqnarray}
where ${\bf F}_{u}^{\mathcal{U}}$ is from (\ref{eq:sumach_zu}).

Correctness (refer to (\ref{corr})) trivially holds as
\begin{eqnarray}
\sum_{u\in\mathcal{U}} X_u^{\mathcal{U}} = \sum_{u\in\mathcal{U}} W_u + \underbrace{ \sum_{u\in\mathcal{U}} {\bf F}_{u}^{\mathcal{U}} \left({\tilde Z}_{u}^{n}\right)^T }_{\overset{(\ref{eq:sumach_zu})}{=} {\bf 0}} \label{eq:sumach_x} = \sum_{u\in\mathcal{U}} W_u
\end{eqnarray}
and we verify the security constraint (\ref{sec}) as follows.
\begin{eqnarray}
&& I \left(\left(W_u\right)_{u \in \mathcal{U}}; \left(X_u^{\mathcal{U}} \right)_{u \in \mathcal{U}} \Bigg| \sum_{u \in \mathcal{U}} W_u\right)  \notag\\
&=& H \left(\left(X_u^{\mathcal{U}} \right)_{u \in \mathcal{U}} \Bigg| \sum_{u \in \mathcal{U}} W_u\right) 
-
H \left(\left(X_u^{\mathcal{U}} \right)_{u \in \mathcal{U}} \Bigg| \sum_{u \in \mathcal{U}} W_u, \left(W_u\right)_{u \in \mathcal{U}}\right)\\
&\overset{(\ref{input_ind}) (\ref{eq:sumach_xu}) (\ref{eq:sumach_x})}{\leq}& n L \log_2 q
- H \left(\left(
{\bf F}_{u}^{\mathcal{U}} \left({\tilde Z}_{u}^{n}\right)^T
\right)_{u \in \mathcal{U}}\right)\\
&\overset{(\ref{eq:sumach_mds})}{=}&
n L \log_2 q - n L \log_2 q = 0 \label{eq:temp}
\end{eqnarray}
where (\ref{eq:temp}) follows from the property that $\left({\tilde Z}_{u}^{n}\right)_{u\in\mathcal{U}}$ are $(K, n)$-MDS.

Finally, when $|\mathcal{U}| = 1$, the problem is trivial as the only selected user may directly send its input to the server. The proof is now complete.

\section{Discussions}
In this work, we characterize the optimal rate of MDS variable generation and secure summation with user selection, somewhat surprisingly and interestingly, as the harmonic number. 

More results can be proved for the settings studied in this work and are summarized below (the proofs are straightforward generalizations of those in \cite{Zhao_Sun_Sum} and this work, thus omitted). For $K$-user MDS variable generation, we may show that the minimum total amount of randomness used in the generation process is $K$ bits for each generated MDS variable bit. Similarly, for $K$-user secure summation with arbitrary user selection, we may show that the minimum total amount of randomness in the keys at all users is $K-1$ symbols for each input symbol. In addition, the protocol of Theorem \ref{theorem2} is also communication--wise optimal, i.e., the minimum message size of each user is $1$ symbol for each input symbol.

While throughout this work, we have highlighted the similarity between MDS variable generation and secure summation with user selection, these two problems might have more than subtle differences. While we have used MDS variable generation in the achievable scheme of secure summation (which indeed turns out to be optimal), we show next that this is not necessary. 
Consider $3$-user secure summation and Table {\ref{table2}} contains an optimal scheme with key rate $R_Z=3/2$ ($L=2, L_Z=3$), which does not rely on MDS variables.
\begin{table}[!ht]
\centering
\begin{minipage}{\textwidth}
\centering
  \begin{minipage}[b]{0.45\textwidth}
   \centering
        \begin{tabular}{ c c c c}
        User 1 & User 2 & User 3 \\ \hline
         $A$ & $A$ & $A$ \\  
         $B$ & $B$ &  \\ 
         $C$ &     &  $C$ \\
             & $D$ &  $D$ 
        \end{tabular}
   \end{minipage}
   \begin{minipage}[b]{0.45\textwidth}
  \centering
       \begin{tabular}{ c| c c c}
        & User 1 & User 2 & User 3 \\ \hline
        \multirow{2}{*}{$(3,1)$-MDS} & $A$ & $A$ & $A$ \\
         & $B$ & $B$ & $B$ \\ \hline
        \multirow{2}{*}{$(3,2)$-MDS}  & $A+B$ & $A+2B$ & $A+3B$\\
        & $C+D$ & $C+2D$ & $C+3D$
         \end{tabular}
  \end{minipage}
\end{minipage}
    \caption{An optimal key design $Z_k$. For comparison, the MDS variables based scheme is shown on the right for prime field $\mathbb{F}_q$, where $q \geq 5$. $A, B, C, D$ are independent and uniform random variables.}\label{table2}
\end{table}

The messages may be set as
\begin{eqnarray}
&& X_1^{\{1,2\}} = W_1 + \left[\begin{array}{cc}
     A\\
     B
\end{array}\right],~
X_2^{\{1,2\}} = W_2 - \left[\begin{array}{cc}
     A\\
     B
\end{array}\right],\notag\\
&& X_1^{\{1,3\}} = W_1 + \left[\begin{array}{cc}
     A\\
     C
\end{array}\right],~
X_3^{\{1,2\}} = W_3 - \left[\begin{array}{cc}
     A\\
     C
\end{array}\right],\notag\\
&& X_2^{\{2,3\}} = W_2 + \left[\begin{array}{cc}
     A\\
     D
\end{array}\right],~
X_3^{\{2,3\}} = W_3 - \left[\begin{array}{cc}
     A\\
     D
\end{array}\right],\notag\\
&& X_1^{\{1,2,3\}} = W_1 + \left[\begin{array}{cc}
     A+B\\
     C
\end{array}\right],~
X_2^{\{1,2,3\}} = W_2 + \left[\begin{array}{cc}
     -B\\
     D
\end{array}\right],~
X_3^{\{1,2,3\}} = W_3 - \left[\begin{array}{cc}
     A\\
     C+D
\end{array}\right]. \notag
\end{eqnarray}
Correctness and security can be easily verified. While this scheme has the same random consumption as the MDS variables based scheme, its keys are uncoded and are thus easier to implement. There could exist further differences if we include additional constraints to the problem. More generally, the difference between MDS variable generation and secure summation may lie in the observation that secure summation only requires $n$ key variables among $n+1$ selected users to be generic (not among all $K$ key variables). Further connections remain to be exploited.

\bibliography{Thesis}

\end{document}